\documentclass[12pt,preprint]{aastex}
\slugcomment{Accepted  for publication  
in {\it  Astronomy and Astrophysics}}  
\def\lax {\ifmmode{_<\atop^{\sim}}\else{${_<\atop^{\sim}}$}\fi}  
\def\gax {\ifmmode{_>\atop^{\sim}}\else{${_>\atop^{\sim}}$}\fi}  
\def\gtorder{\mathrel{\raise.3ex\hbox{$>$}\mkern-14mu
             \lower0.6ex\hbox{$\sim$}}}

\def\cm2{cm$^{-2}$}
\def\s1{s$^{-1}$}

\begin{document}


\title{BL Lacertae: X-ray spectral evolution and a black-hole mass estimate
}




\author{Lev Titarchuk\altaffilmark{1} \& Elena Seifina\altaffilmark{2}}
\altaffiltext{1}{Dipartimento di Fisica, Universit\`a di Ferrara, Via Saragat 1, I-44122 Ferrara, 
Italy, email:titarchuk@fe.infn.it; 
National Research Nuclear University MEPhI (Moscow Engineering Physics Institute), Moscow, Russia}
\altaffiltext{2}{Moscow State University/Sternberg Astronomical Institute, Universitetsky 
Prospect 13, Moscow, 119992, Russia; seif@sai.msu.ru}

\begin{abstract}
{We present an analysis of the spectral properties observed in X-rays from active galactic nucleus 
BL Lacertae using {\it RXTE}, {\it Suzaku}, {\it ASCA}, {\it Beppo}SAX, and {\it Swift} observations. 
The total time covered by these observations  is approximately 20 years. We show strong observational evidence 
that  this source undergoes X-ray spectral transitions from the low hard state (LHS) through the intermediate 
state (IS) to the high soft state (HSS)  during these observations. During the {\it RXTE} observations 
(1997 -- 2001, {180 ks, for a total 145 datasets}), the source was approximately $\sim 75$\%, 
$\sim 20$\% and only $\sim 5$\% of the time  in the IS, LHS, and HSS, respectively. We also used  $Swift$ 
observations (470 datasets, for a total 800 ks), which occurred during 12 years (2005 -- 2016), the 
broadband (0.3 -- 200 keV) data  of $Beppo$SAX (1997 -- 2000, 160 ks),  and the low X-ray energy (0.3 -- 10 keV) 
data of $ASCA$ (1995 -- 1999, 160 ks). Two observations of $Suzaku$ (2006, 2013; 50 ks) in combinations  
with long-term $RXTE$ and $Swift$ data-sets fortunately  allow us to describe  all spectral states of BL~Lac. 
The spectra of BL Lac are well fitted by the so-{called} bulk motion Comptonization (BMC) model  for all spectral states. 
We have established the photon index  saturation level, $\Gamma_{sat}$=2.2$\pm$0.1, in the $\Gamma$ versus  mass 
accretion rate ($\dot M$) correlation. This $\Gamma-\dot M$ correlation allows us to estimate the black-hole (BH) 
mass in BL Lac to be $M_{BH}\sim 3\times 10^7 M_{\odot}$ for a distance of 300 Mpc. For the BH mass estimate, 
we use the scaling method taking stellar-mass Galactic BHs  4U~1543--47 and GX~339--4 as reference sources. 
The $\Gamma-\dot M$ correlation revealed in BL Lac  is similar to those in a number of stellar-mass Galactic BHs 
and two recently studied intermediate-mass extragalactic BHs. It  clearly shows the correlation along with 
the very extended $\Gamma$ saturation at  $\sim 2.2$. This is   {robust} 
observational evidence for the presence of a BH in BL Lac. We also reveal that  the  seed (disk) photon 
temperatures are relatively low, of order of 100 eV, which are consistent with  a high BH mass in BL Lac. 
It is  worthwhile to emphasize that we found particular events when X-ray emission anti-correlates with radio 
emission. This  effect  indicates that mass accretion rate (and thus  X-ray radiation) is higher when the mass 
outflow is lower.  
} 
\end{abstract}

\keywords{accretion, accretion disks --
                black hole physics --
                stars, galaxies: active -- galaxies: BL Lacertae objects: general -- galaxies: BL Lacertae objects: 
individual: BL Lacertae --
                radiation mechanisms 
}

\section{Introduction}

BL Lacertae (BL Lac),  B2200+420 is a highly variable  active galactic nucleus (AGN),
which  was discovered  approximately a century  ago by Cuno Hoffmeister~ [see \cite{Hoffmeister29}]. This source 
 was initially thought to be simply an irregular variable star in the Milky Way galaxy and for this reason 
was given a variable star notation. Later, the radio counterpart was identified for this ``star'' 
by John Schmitt at the David Dunlap Observatory [\cite{Schmitt68}]. 
When Oke and Gunn (1974) measured the redshift of BL Lac ($z=0.069$) it  became clear that this object is located
at a distance of 900 million light years (or $\sim$ 300 Mpc). BL Lac is 
{
the eponym for BL Lacertae objects. 
}
This class is distinguished by optical spectra where   broad emission  lines  are absent. Note,  the broad emission lines   are usually characteristics of quasars.   Nonetheless,BL Lac
  sometimes displays weak emission lines. 
For this reason, BL Lacs  are classified based on overall spectrum. Specifically, the BL Lac objects  with a low-energy peak located in the UV or X-rays and usually  found during  X-ray surveys  were labeled as ``high-energy peaked BL Lacs'' or HBLs (see Giommi, Ansari, \& Micol 1995; Madejski et al. 1999). While those with the lower-energy peak 
in the  infra-red (IR) range  were defined as ``low-energy peaked BL Lacs'' or LBLs. 

To understand blazar variability we should study  wide band spectra during major flaring episodes 
and BL Lac has been a target of many  such multi-wavelength  campaigns (see  for example, 
Ravasio et al. 2003). 

%
%
 \begin{figure*}
 \centering
\includegraphics[scale=0.8,angle=0]{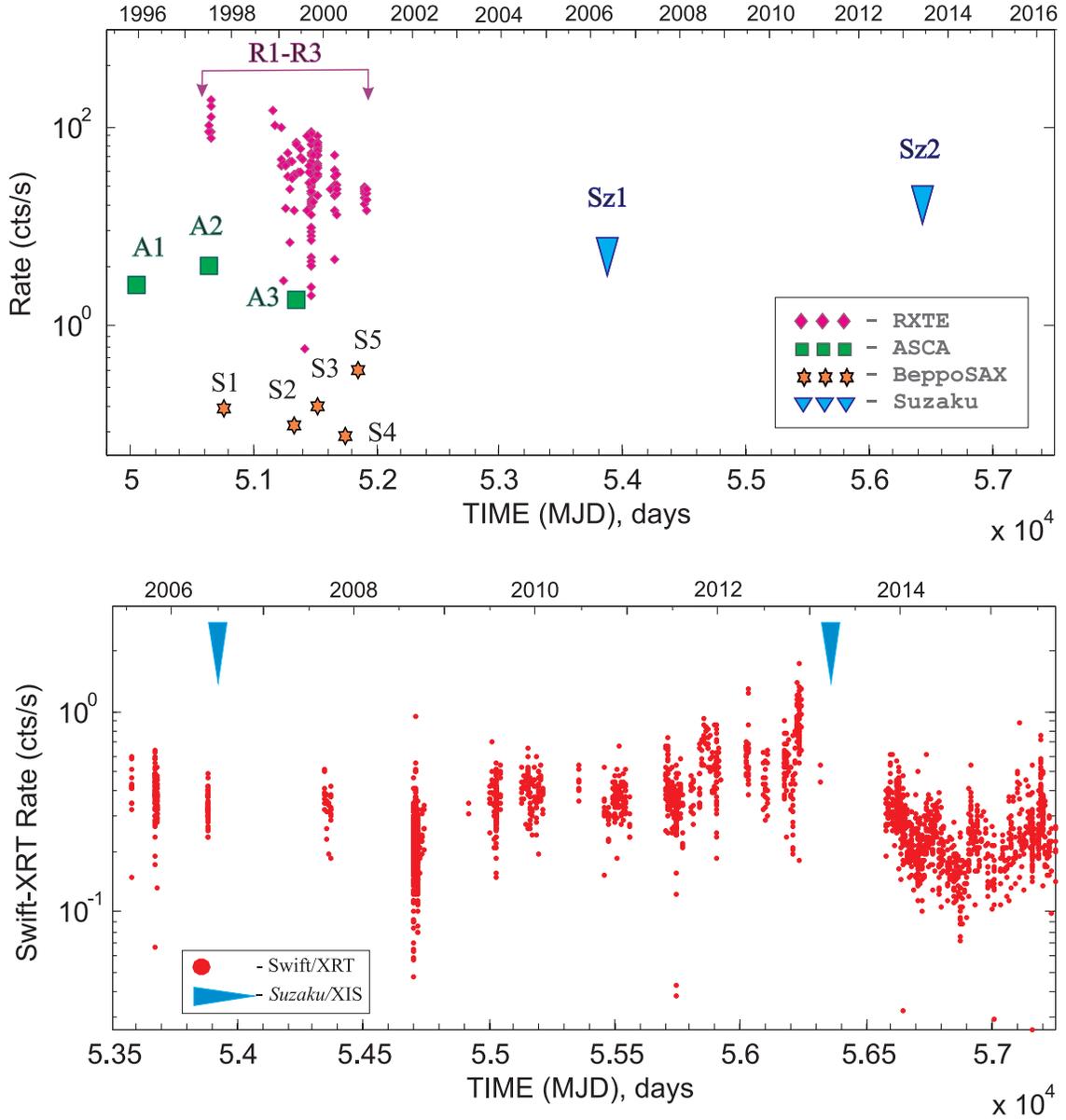}
     \caption{
{ 
{\it Top}: The time distribution of $ASCA$ (green squares, ``A''-marks), $RXTE$ (pink diamonds, ``R''-marks), 
$Beppo$SAX (brown stars, ``S''-marks), and $Suzaku$ (blue triangles, ``Sz''-marks) observations 
(see Tables~\ref{tab:list_RXTE} -- \ref{tab:list_SAX_Suzaku}). 
{\it Bottom}: The $Swift$/XRT light curve of BL Lac in the 0.3$-$10 keV  range during 2005 -- 2016.  Red points mark the source signal (with 2-$\sigma$ detection level) and 
blue arrows  show the MJD of $Suzaku$.
Note, that rate-axis is related to {\it RTXE}/PCA count rate which 
are not comparable with  other instruments ({\it ASCA, BeppoSAX} and {\it Suzaku}). For clarity, the error bars are omitted.
}
}

   \label{lc}
 \end{figure*}
BL Lac is well known for its prominent variability in a wide  energy range,  particularly, its variability  in optical (Larionov et al., 2010; Gaur et al., 2015) and radio (Wehrle et al., 2016).
Raiteri et al. (2009) showed  the broadband observations from radio to X-rays of BL Lac, which were taken  during the 2007 -- 2008   Whole Earth Blazar Telescope (WEBT) campaign.  They fitted the spectra  
by an inhomogeneous, rotating helical jet model, 
which includes synchrotron self-Compton (SSC) emission from a helical jet plus a thermal component
from the accretion disk (see also Villata \& Raiteri 1999; Ostorero et al. 2004; Raiteri et al. 2003). Larionov et al. (2010) studied the behavior of BL Lac optical flux and color variability 
and suggested the variability to be mostly caused by changes of the jet. Raiteri et al. (2010) investigated  the broad band emission and timing properties of  BL Lac during the  2008 -- 2009 period and argued  for a jet geometry model  where changes in the  viewing angle of the jet emission regions played an important role in the source's multiwavelength behavior. Moreover,  Raiteri et al. (2013) collected  an extensive optical sampling using the GLAST-AGILE Support Program of the WEBT for the BL Lac outburst period  during 2008 -- 2011 and  tested cross-correlations between the optical-$\gamma$-ray and X-ray-mm bands. 

BL Lac has been observed by many missions in the  X-ray energy band and surveys were conducted 
using satellites 
HEAO--1, {\it Einstein}, {\it Ginga, ROSAT, ASCA, BeppoSAX, RXTE,} and {\it Swift} during a 20-year period.
Thereafter, the total number of observations and total exposure times of the {\it RXTE} and {\it Swift} observations (and  these two datasets give the most extensive time coverage, the latter covering 2005-2016) and  a further three missions covered only a limited number (from two to five) of epochs but with the largest effective area (i.e., best spectral signal-to-noise ratio  (SNR) spectra).
In the earliest data, Bregman et al. (1990),  for example, using the {\it Einstein} observations found 
{the best-fit} values of   $\Gamma$  to be approximately    1.7, on average,  while Urry et al. (1996),  using $ROSAT$  data, estimated   
$\Gamma$ to be 1.95$\pm$0.45.  Using $GINGA$ observations Kawai et al. (1991)  found $\Gamma\sim$  
1.7 -- 2.2. 

X-ray variability of BL Lac is  less well studied in terms of spectral state transition, which is widely investigated  in Galactic sources (see e.g., Shaposhnikov \& Titarchuk 2009, hereafter ST09).  
Raiteri et al. (2010) (see Figure 9 therein), constructed spectral energy distributions (SEDs) of BL Lac corresponding to two  epochs, when the source had different brightness levels: 2008 (May combined with August) and 1997 (July) based on contemporaneous data from $Swift$ (UV and X-ray), GASP-WEBT
(optical and radio) and $Fermi$ ($\gamma$-ray data, Abdo et al. 2010). 
These SEDs have two strong peaks which  are usually associated with the synchrotron and
SSC components and they 
 vary in amplitude, spectral shape  and  peak frequencies. 
In this paper we  concentrate our efforts  on studying X-ray variability for BL Lac in the energy range from 0.3 to 150 keV.

It is usually believed that BL Lac contains a supermassive black hole (SMBH) but there is  no direct estimate of the value of its BH mass.  Magorrian et al. (1998) and Bentz et al. (2009), constructing dynamical models,  established correlations between the luminosity of the bulge of the host galaxy and a BH  mass, and Ferrarese \& Merritt (2000); G$\ddot u$ltekin et al. 
(2009)  found  a correlation between the velocity  dispersion  and a BH mass. 
Moreover,  Kaspi et al. (2000),  Vestergaard (2002), and   Decarli et al. (2010) established   the correlation between the luminosity of the continuum at selected frequencies and the size of the Broad Line Region (BLR). They suggested using this correlation  to estimate BH mass.  In  particular,  for cases of very powerful blazars, where   their IR-optical-UV continuum is dominated 
by a thermal component, these authors suggested that  the thermal component is related to  the accretion disk
(Ghisellini et al. 2011). Thus, modeling of this radiation using  the standard Shakura-Sunyaev  disk allows one to estimate  a black hole mass as well as  an accretion rate.
 
More details of the methods of  BH mass estimates in AGNs can also be revealed from studies based on, for example,  reverberation mapping, kinematics in the bulge of the host galaxy of AGN (\cite{bm82}; Peterson (1993), (2014); Peterson et al. (2014); Ferrarese \& Merritt (2000); Ryle 2008;  G$\ddot u$ltekin et al. (2009)), and break frequency scales
(see  Ryle (2008)). 
A high BH mass value of its central source in BL Lac actually can give high luminosity  along with variability of its emission in  all energy bands. 
For a BH mass estimate in BL Lac, 
Liang \& Liu (2003) used lower  timescales of variability, although these estimates are  not a robust indicator of a BH mass. Nevertheless, the authors  relate them with  the photon  light-crossing time.  Mass determination using this particular method gives  a black hole mass, $M_{BH} \sim  3\times 10^{6}$ M$_{\odot}$ for BL Lac 
which differs from  that determined using emission of the fundamental plane technique (see Woo \& Urry  2002). Using this   
method,  
Woo \& Urry   found that $M_{BH} = 1.7 \times 10^{8}$ M$_{\odot}$ for BL Lac.

Therefore, it is desirable to have an independent BH identification  for its central 
object as well as the BH mass determination  by an alternative to the abovementioned methods, 
based on luminosity estimates only. 
A method of  BH mass determination was developed by Shaposhnikov \& Titarchuk (2009), hereafter ST09, using a correlation scaling between X-ray spectral and timing (or mass accretion rate); properties observed for many Galactic BH binaries during  
their spectral  state transitions. 

We apply the ST09 method to {\it RXTE}, {\it Beppo}SAX, {\it ASCA}, {\it Suzaku,} and  {\it Swift}/XRT
  data of BL Lac.  Whether or not the observed spectral variability  of BL Lac  can be explained in terms of spectral state transition remains to be seen. We fitted the X-ray data applying  the bulk motion Comptonization (BMC) model along with photoelectric absorption.  
The parameters of the BMC model are the seed photon  temperature $T_s$, the energy index of the Comptonization spectrum  $\alpha$ ($\alpha=\Gamma-1$), and the illumination parameter $\log(A)$ related to the Comptonized (illumination) fraction $f=A/(1+A)$. This model uses a convolution of  a seed  blackbody  with an upscattering   Green's function, presented in the framework of the BMC as a broken power law whose left and right wings have indices $\alpha+3$ and $\alpha$, respectively (we refer to the description of the BMC Comptonization Green's function in Titarchuk  \& Zannias (1998) (TZ98) and suggest comparison with Sunyaev \& Titarchuk (1980)). 

Previously, many properties of BL Lac were  analyzed using {\it Swift}/XRT observations. 
In particular, Raiteri et al. (2010) analyzed the {\it Swift} (2008 -- 2009) observations  and later Raiteri et al. (2013) investigated 
the {\it Swift} observations made from 2009 to 2012 (see the light curve in Figure~\ref{lc}),  using fits 
of their X-ray spectra with a simple absorbed power law model. 
They found that in the X-ray spectra of BL Lac, the values of $\Gamma$  are scattered between 1.32 (hard spectrum) and 2.37 (soft spectrum) without  any correlation with the flux.

 {\it ASCA} observed BL Lac in 1995 detecting the photon index, $\Gamma$  $\sim1.94$ 
(Madejski et al. 1999; Sambruna et al. 1999). {\it RXTE} found a harder spectrum with $\Gamma$  in the range 
1.4 -- 1.6 over a time span of seven days (Madejski et al. 1999). A fit of simultaneous {\it ASCA} and 
{\it RXTE} data shows the 
existence of a very steep and varying soft component for the photon energies $E<$ 1 keV. Photo index  $\Gamma$    was in the  range 3 -- 5, in addition to the hard power law component with $\Gamma$  in the range 1.2 -- 1.4. Two rapid flares with 
time scales of 2 -- 3 hours were detected by {\it ASCA} 
in the soft part of the spectrum (Tanihata et al. 2000). In November 1997, BL Lac was observed using the $Beppo$SAX, see Padovani et al. (2001), who estimated $\Gamma$ in the interval of 1.89$\pm$0.12 .

In this paper we present an analysis of  available {\it Swift}, {\it Suzaku}, {\it Beppo}SAX, {\it ASCA} and {\it RXTE} observations 
of BL Lac 
in order to re-examine previous 
conclusions on a BH 
as well as to find further indications to a supermassive BH in BL Lac.  
In \S 2 we show  the list of observations used in our data analysis while 
in \S 3 we provide  details of the X-ray spectral analysis.  We discuss an evolution of 
the X-ray spectral properties during the high-low state  transitions 
and demonstrate the results of the scaling analysis in order to estimate a BH mass of BL Lac in \S 4. 
We  make  our conclusions on the results  in \S 5. 

\section{Observations and data reduction \label{data}}

{
We examined X-ray data of BL~Lac using a number of instruments with various spectral capabilities covering different energy 
ranges. 
{\it Beppo}SAX and {\it RXTE} cover very wide energy ranges 
from 0.3 to 200 keV and from 3 keV to 100 keV, respectively.  In Sect.~\ref{rxte data} 
we present our analysis of the {\it RXTE} archival data  from 1997 to 2001 (for our sample of 145 observations; approximately 180 ks). 
We also investigate the {\it Beppo}SAX archival data  taken around   the time  interval of the $RXTE$ observations
(see Sect.~\ref{sax data}). }

 In the {upper} panel of Figure~\ref{lc} we show  the time distribution of $RXTE$ and  $Beppo$SAX indicated by  $pink$ diamonds  with ``$R$''-marks and  
 $brown$ stars with ``$S$''-marks, respectively.
{It is worth noting   that the  count rates of  individual satellites are not comparable between each other. 
 {\it Beppo}SAX observations are not so numerous (only five observations) while they cover all spectral states of BL~Lac 
with high energy sensitivity and long time exposure (for a total  of 160 ks). 

{Along with these high-energy 
observations we also investigate  spectral evolution of BL Lac during long-term observations  
(2005 -- 2016)  by the $Swift$ X-ray Telescope 
(XRT; Burrows et al. 2005) in lower (0.3 -- 10 keV) energy band (see details in Sect.~\ref{swift data})}. 
{$Swift$  data contain over {470} detections of BL~Lac with a total exposure of 800 ks over a 12 year interval (see bottom panel 
of Fig.~\ref{lc})}. 
{We also use for our analysis three observations by  $ASCA$ with total exposure of  160 ks (1995 -- 1999; see Sect.~\ref{asca data}),  and two detections by $Suzaku$  with  an exposure time  of approximately 50 ks (2006, 2013; see 
Sect.~\ref{suzaku data})}. {The MJDs for two $Suzaku$ observations of BL~Lac are indicated by blue arrows  in 
the light curve of BL~Lac obtained by $Swift$ and shown in the bottom panel of Figure~\ref{lc}}. {Note,
$Suzaku$ points are  present  in  the {upper} panel to compare with $RXTE$, $ASCA,$ and $Beppo$SAX observations.

{All these datasets  are spread over twenty years  (see Figure~\ref{lc}), sometimes randomly and sometimes covering common intervals.  $Suzaku$ provides much better photon statistics 
than  $Swift$ due to the long $Suzaku$ exposures. The total list of BL~Lac observations used in our analysis is
given in Tables~1$-$5.
We applied the nominal position of BL~Lac ($\alpha=22^{h}02^{m}43^s.3$, $\delta=+42^{\circ} 16{\tt '} 39{\tt ''}$,  J2000.0 (see, e.g., Ghisellini et al., 2011)\footnote{http://deepspace.jpl.nasa.gov/dsndocs/810-005/107/catalog-fixed.txt}.

  \begin{figure*}[ht]
 \centering
    \includegraphics[width=17cm]{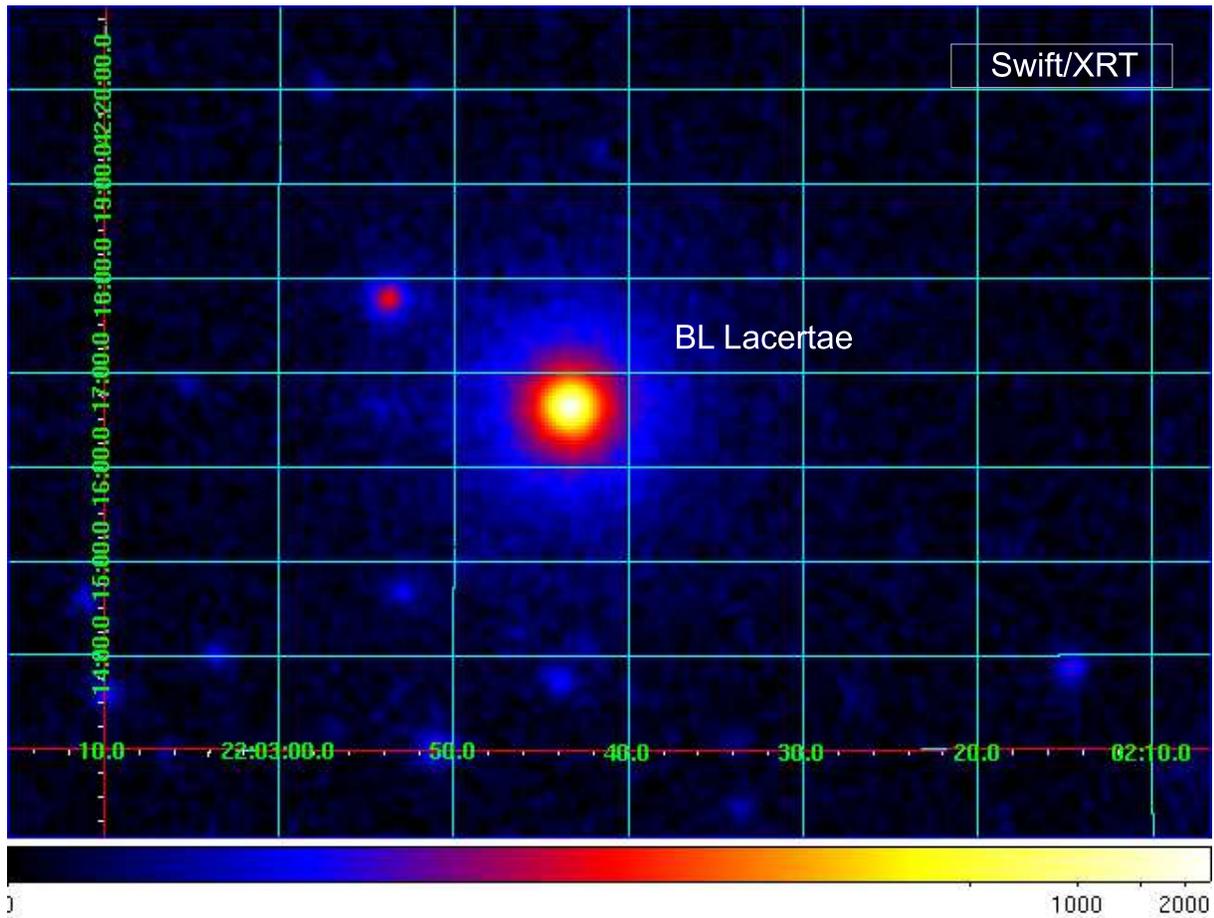}
      \caption{
The $Swift$/XRT (0.3 -- 10 keV) image  of the BL~Lac field taken during 2005 -- 2016 (800 ks) 
{ 
centered on the nominal position of BL~Lac ($\alpha=22^{h}02^{m}43^s.3$, $\delta=+42^{\circ} 
16{\tt '} 39{\tt ''}$,  J2000.0). The field is approximately $9^{\prime}\times 15^{\prime}$. 
}
}
\label{image}
\end{figure*}

 \begin{figure*}
 \centering
\includegraphics[width=13cm]{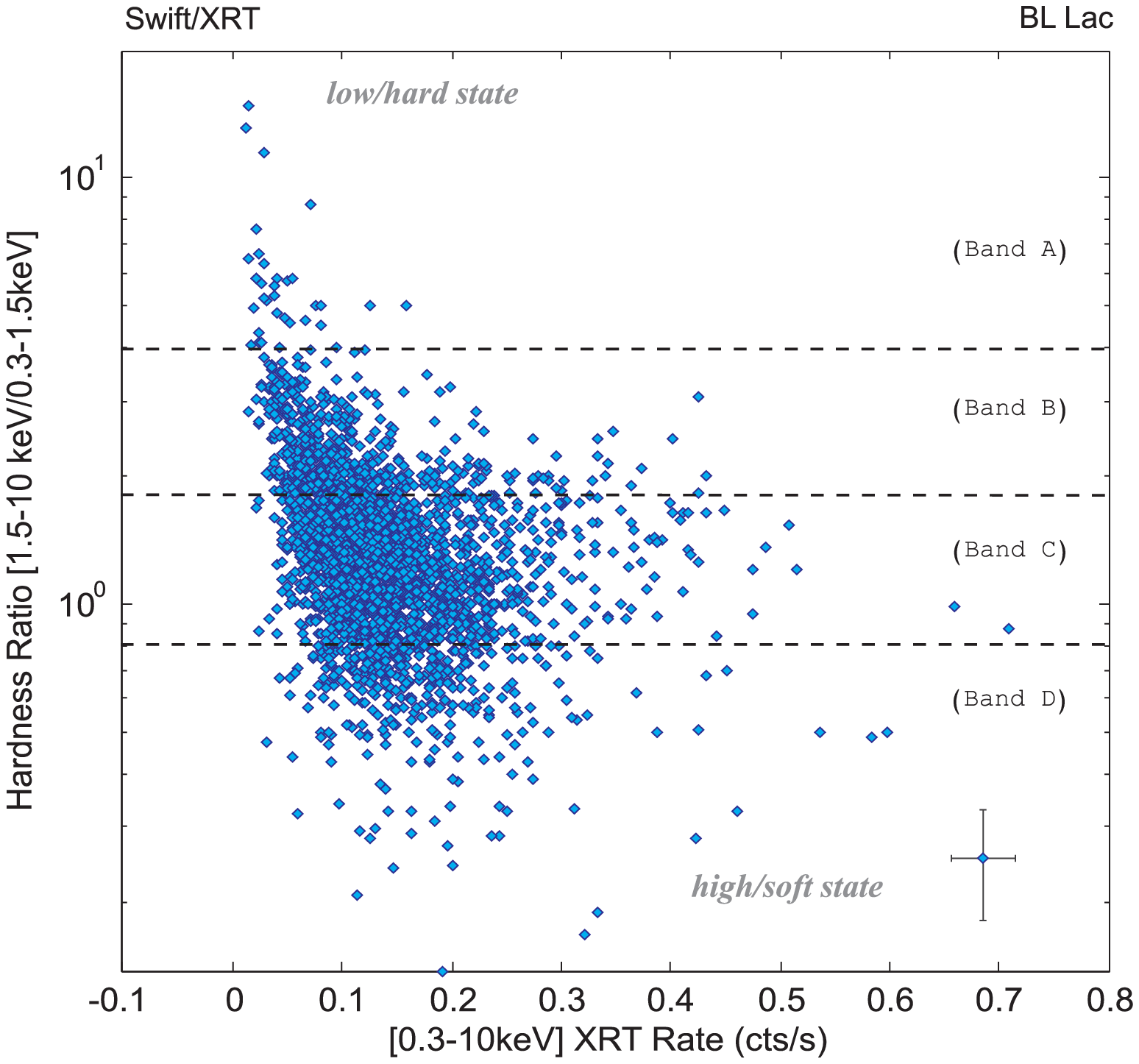}
   \caption{
Hardness-intensity 
diagram for BL Lac using {\it Swift} observations (2005 -- 2016)  
during spectral evolution from the low/hard state to the high/soft states. 
 In the vertical axis, the hardness ratio (HR)  is a ratio 
of the source counts in the  two energy bands: 
the {hard} (1.5 -- 10 keV) and  soft (0.3 -- 1.5 keV). 
The HR  decreases with a source  brightness in the 0.3 -- 10 keV  range (horizontal axis). 
{ For clarity, we plot only one point with error bars (in the bottom right corner) to demonstrate 
typical uncertainties  for the count rate and HR.} 
   \label{HID}
}
 \end{figure*}

We extracted all of these data from the HEASARC archives and found that they 
encompassed a wide range of X-ray luminosities. 
The well-exposed {\it ASCA}, {\it Beppo}SAX, and {\it Suzaku} data are affected  by the low-energy photoelectric 
absorption, which is presumably not related to the source. The fitting was carried out using the standard {\tt XSPEC} astrophysical fitting package.
}


\subsection{\it RXTE \label{rxte data}}

For our analysis, we  used {145} {\it RXTE} observations  taken between July 1997 and January 2001 
related to different spectral states of the source. 
Standard tasks of the LHEASOFT/FTOOLS 5.3 software package were 
applied for data processing. For spectral analysis we 
used PCA {\it Standard 2} mode data, collected in the 3 -- 23~keV energy range, 
using PCA response calibration (ftool pcarmf v11.7). The standard dead time correction procedures were applied to the data. 
In order to construct broad-band spectra, the data from HEXTE detectors were also used.
 The spectral analysis of  the data  in the 19 -- 150~keV energy range should also
be implemented  in order 
to account for the uncertainties in the HEXTE response and 
background determination.
We  subtracted the background corrected  in  off-source observations. 
 The data { of BL~Lac} are available through the GSFC public archive 
(http://heasarc.gsfc.nasa.gov). Systematic error of 0.5\% has been applied to the derived spectral parameters  
of  {\it RXTE}  spectra. 
In  Table~\ref{tab:list_RXTE} we list the  groups  of {\it RXTE} observations tracing 
the source evolution during different states. 


\subsection{{\it Beppo}SAX \label{sax data}}

We used the {\it Beppo}SAX data 
of BL Lac carried out from 1997 to 2000 and found the source was in different spectral states. 
In Table~\ref{tab:list_SAX} we show the summary of the {\it Beppo}SAX 
observations analyzed  in this paper. 
Broad band energy spectra of BL Lac were obtained combining data from  three 
{\it Beppo}SAX Narrow
Field Instruments (NFIs): the Low Energy Concentrator
Spectrometer (LECS) for the 0.3 -- 4 keV range (\citet{Parmar97}), the Medium Energy Concentrator Spectrometer
(MECS)  for the 1.8 -- 10 keV range (\citet{boel97}),  and the Phoswich Detection
System (PDS)  for the 15 -- 200 keV range (\citet{fron97}). 
The SAXDAS data analysis package is used for the data processing. 
We performed a spectral analysis for each of the instruments in a corresponding 
energy range within 
which a response matrix is well specified. 
The spectra were accumulated 
using an extraction region of 8 and 4 arcmin radius for the LECS and MECS, respectively. 
The LECS data were renormalized to match the MECS data. Relative normalizations of the NFIs were treated 
as free parameters in the
 model fits, except for the MECS normalization that was fixed at unity.  
 The obtained cross-calibration factor  was found to be in a standard range for each
 instrument
\footnote{http://heasarc.nasa.gov/docs/sax/abc/saxabc/saxabc.html}.
 Specifically, LECS/MECS re-normalization ratio is 0.72 and PDS/MECS
 re-normalization ratio is 0.93. 
While the source was bright and the background was low and stable, we checked its uniform distibution 
across the detectors. Furthermore, we extracted a light curve from a source-free region far from source and 
found that  the background did not vary for the whole observation. 
In addition, 
the spectra were rebinned  in accordance with the energy resolution of the instruments 
using rebinning template files in GRPPHA of
 XSPEC\footnote{http://heasarc.gsfc.nasa.gov/FTP/sax/cal/responses/grouping} 
to obtain better signal to noise ratio  
  for  derivation of the model  spectral parameters.  
We applied systematic uncertainties  of 1\%  to the derived spectral parameters  of  {\it Beppo}SAX  spectra. 


\subsection{\it ASCA \label{asca data}}

$ASCA$ observed BL Lac on  November 22, 1995, on  July 18, 1997 and June 28, 1999. 
Table~\ref{tab:list_asca} summarizes the start time, end time, and the MJD interval  for each of these observations.
For the {\it ASCA} description,  see Tanaka, Inoue, \& Holt (1994).
The {\it ASCA} data were screened using the ftool ascascreen and the standard screening criteria. The
pulse-height data for the source were extracted using spatial regions with a diameter of 3' (for SISs) and 4' (for GISs)
centered on the nominal position of BL Lac, 
while the background was extracted from source-free regions of comparable size away from the source. The spectrum data 
were rebinned to provide at least 20 counts per spectral bin in order to validate  using
 the $\chi^2$ statistic. The SIS and GIS data were fitted applying {\tt
XSPEC} in the energy ranges 0.6 -- 10 keV and 0.7 -- 10 keV respectively, where the spectral responses
are well  known. 


\subsection{\it Suzaku \label{suzaku data}}

For the {\it Suzaku} data  (see Table \ref{tab:list_SAX_Suzaku}) we used the  {\tt HEASOFT software package} (version 6.13) and calibration database (CALDB) released on  February 10, 2012.
 We applied the unfiltered event files for each of the operational XIS detectors (XIS0, 1 and 3) 
and following  the {\it Suzaku} Data Reduction Guide\footnote{http://heasarc.gsfc.nasa.gov/docs/suzaku/analysis/}. 
 We obtained cleaned event files by re-running the {\it Suzaku} {\tt pipeline} implementing the
latest calibration database (CALDB) available
since January 20, 2013,  and also apply the associated screening criteria files. 

Thus, we obtained the BL Lac spectra  from the filtered XIS event data 
taking a circular region, centered on the source, of radius 6{\tt '}. 
Using the {\it Beppo}SAX sample, we considered the 
background region to be in the vicinity  of the source   extraction region. 
We  obtained  the spectra and light curves   from the cleaned event
files using XSELECT, and we generated  responses  for each
detector utilizing  the XISRESP script with a medium resolution.
The spectra and response files for the front-illuminated
detectors (XIS0, 1 and 3) were combined using the FTOOL
ADDASCASPEC, after confirmation of  their consistency. 
Finally, we again grouped the spectra to have a minimum of 20 counts per energy bin.

We carried out  spectral fitting  applying XSPEC package. 
 The energy ranges at approximately 1.75 and 2.23 keV were not  used for spectral fitting because of the known  artificial structures in the XIS spectra around the Si and Au edges.
 Therefore, for spectral fits we  took  the 0.3 -- 10 keV  range  for the XISs 
(excluding 1.75 and 2.23 keV points). 


\subsection{{\it Swift} \label{swift data}}

Since the effective area of the $Swift$/XRT is less than 
for the $Suzaku$/XIS and $Beppo$SAX detectors in the
0.4 -- 10 keV range, detailed spectral modeling is difficult to make using $Swift$ data only.
Therefore, we analyzed the XRT data in the framework of  the BMC model and used  photoelectric absorption determined  by $Beppo$SAX spectral analysis.  

%

We used $Swift$ data carried out from 2005 to 2016. 
In Table~\ref{tab:list_Swift} we show the summary of the {\it Swift}/XRT 
observations analyzed in this paper. In the presented $Swift$ observations, BL Lac shows global outburst peaked in 
 November 2012 (see Figure~\ref{lc}, {bottom} panel) as well as moderate variability and low flux level intervals, when 
the source has been detected  at least, at $\sim$ 2-$\sigma$ significance (see, Evans et al. 2009). 
The $Swift$-XRT data in photon counting (PC) mode 
(ObsIDs, indicated in the {second} 
column of 
Table~\ref{tab:list_Swift}) were processed using the HEA-SOFT v6.14, the tool XRTPIPELINE v0.12.84, and the
calibration files (latest CALDB version is 20150721\footnote{http://heasarc.gsfc.nasa.gov/docs/heasarc/caldb/swift/}).
The ancillary response files were created using XRTMKARF v0.6.0 and exposure maps generated by XRTEXPOMAP v0.2.7. 
We fitted the spectrum using the response file SWXPC\-0TO12S6$\_$20010101v012.RMF.
We also applied the online XRT data product generator\footnote{http://www.swift.ac.uk/user\_objects/}  for independent verification of  
light curves and spectra (including background and ancillary response files, 
see Evans et al. 2007, 2009). 


\section{Results \label{results}}

\subsection{Images \label{image_lc}}

We made a visual inspection of the source field of view (FOV) image  to get rid of a possible contamination 
from nearby sources. {The Swift/XRT (0.3 -- 10 keV) image of BL~Lac FOV is shown
in Fig.~\ref{image}. It is evident that while  
some sources are presented in BL~Lac FOV, they are far from BL~Lac (seen clearly in the center of the FOV). 
Thus, we excluded the contamination by other bright point sources within a 10 arcmunute 
radius circle.
}

\subsection{Hardness-intensity diagrams and light curves  \label{HID_lc}}

Before we proceed with 
details of the  spectral fitting  we  study 
a hardness ratio (HR) { as a function of soft counts in the  0.3 -- 1.5 keV band} using    
 the {\it Swift} data. Specifically, we consider the  {HR} as a  ratio of the hard counts 
(in the 1.5 -- 10 keV range) and the soft ones.} 
The HR is evaluated 
by careful calculation of  background counting{ and uses only significant points (with 5-$\sigma$ detection)}.
In Figure~\ref{HID}   we demonstrate  
the  hardness-intensity diagram (HID)  and thus, we show  that different count-rate observations 
 are associated with 
 different color regimes. Namely, the HR larger values correspond to  harder spectra.
A Bayesian approach was used  to estimate   the HR values and their errors~[\citet{Park06}]\footnote{A Fortran and C-based program which calculates the 
ratios using the methods described by \cite{Park06} 
(see http://hea-www.harvard.edu/AstroStat/BEHR/)}.

%
%

\begin{figure*}
\begin{center}
   \resizebox{\hsize}{!}{\includegraphics{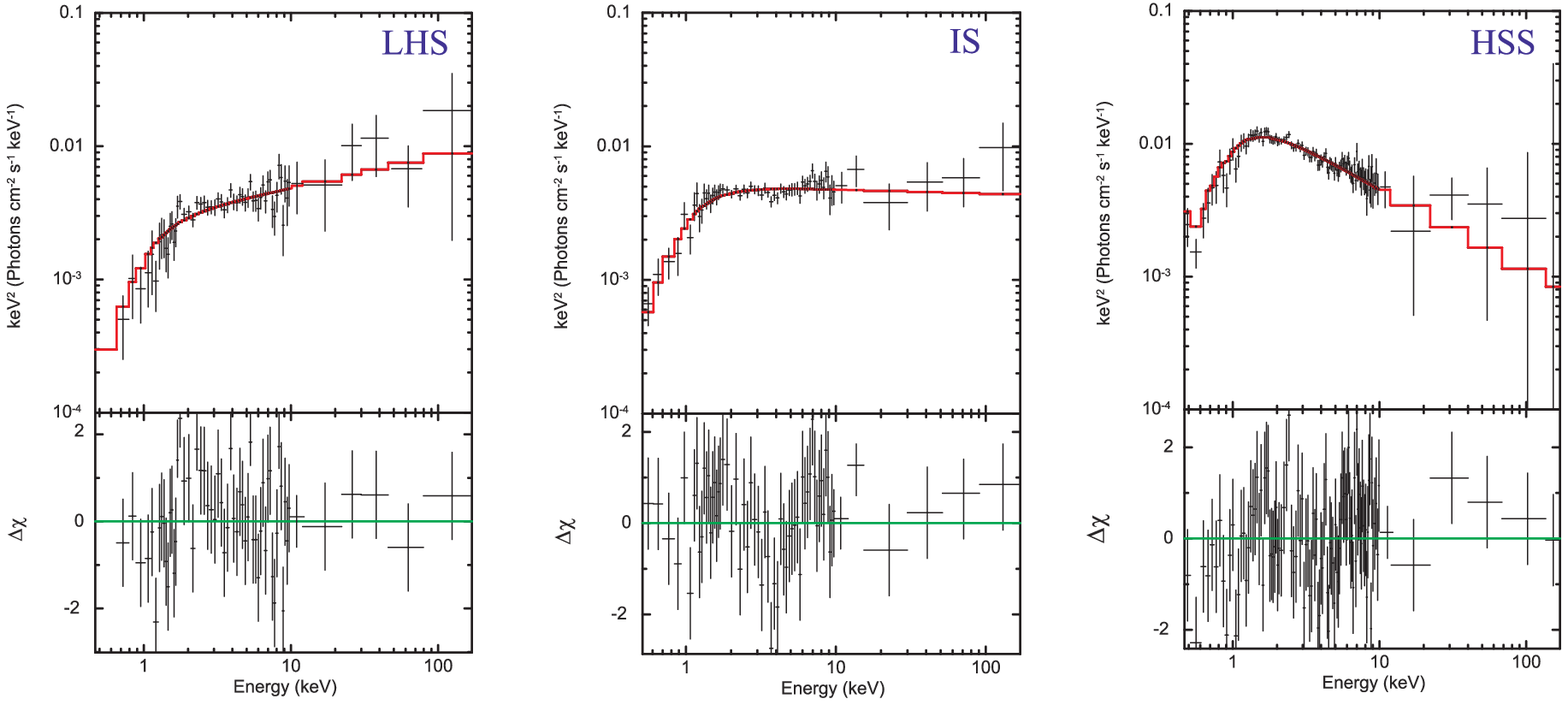}}
\end{center}
  \caption{
Three representative $EF_E$  diagrams for different states of BL Lac. 
Data are taken from $Beppo$SAX observations { S1} (left panel, LHS), 
{S2} ({central} panel, IS), 
and { S5} (right panel, HSS). 
The data are shown by black crosses and  
 the spectral model {(phabs*BMC)} is displayed  as a red line. 
}
\label{BeppoSAX_spectra}
\end{figure*}

Figure~\ref{HID}   
indicates that the {HR} monotonically reduces  with the total count rate (in the 0.3 -- 10 keV energy band). 
This particular sample  is similar to those of  most outbursts of Galactic 
X-ray binary transients (see Belloni et al. 2006; Homan et al. 2001; Shaposhnikov \& Titarchuk, 2006;  ST09; TS09; Shrader et al. 2010; Mu$\tilde n$oz-Darias et al. 2014).

 We show the {\it Swift}/XRT light curve of BL Lac from 2005 to 2016 for the 0.3 -- 10 keV band  in Figure~\ref{lc}. 
{Red} points mark the source signal. 
{Thus, one can see that BL~Lac shows rapid variability on timescales of less than 10 ks, while}  
in 2012 the source showed{
a  higher  count level, which could be  associated with} 
the global outburst. The   maximum  of the outburst was  
at approximately
MJD 56250
{
for a total rise-decay sample of 2.5 years.
} 
For{ most} 
of the {\it Swift} observations the source remained in the{ low or intermediate state and was in 
the soft state for  only 5\% of the time. 
We should point out that individual {\it Swift}/XRT observations of BL Lac in PC 
mode do not have enough counts into make  statistically significant spectral fits. 

{
Based on the hardness-intensity diagram for BL~Lac (see Fig. \ref{HID}) we also  made the state identification using the hardness 
ratio. 
This plot 
indicates a continuous distribution of the HR with source intensity from
high hardness ratio at lower count-rate to low hardness ratio at higher count events. 
Furthermore, the hardness$-$intensity diagram shows a smooth track. Therefore, 
we  grouped the $Swift$ spectra into four bands according to count rates:
} 
very high ("A", HR>4), high ("B", $1.9<HR<4$), intermediate ("C", $0.8<HR<1.9$), and  
low ("D", $HR<0.8$) count rates to resolve this problem. 
In addition, all  groups of the {\it Swift} spectra were binned to a minimum of
20 counts per bin in order to use $\chi^2$-statistics for our spectral fitting.
Thus, we  combined 
the spectra  in each  related band, regrouping  them with the task grppha and then
we fitted 
them using the 0.3 -- 10 keV  range. 

\subsection{X-Ray spectral analysis \label{spectral analysis}}

{
Various  spectral models were used  in order to test them  for all available data sets for BL Lac.
We wanted to establish  the low/hard and high/soft state evolution using   spectral modeling.
We investigate the {\it Suzaku}, {\it $Beppo$SAX, ASCA, RXTE}, and combined {\it Swift} spectra 
to check  the following spectral models: 
powerlaw, blackbody, BMC and their possible combinations modified by an absorption  model. 

Since BL Lac is located relatively close to the Galactic plane ($b=-10.43$ deg),
it is necessary to take into account the possible contribution from the
Galactic molecular gas in addition to that associated with neutral hydrogen
21 cm values. Furthermore, the source is located behind a molecular cloud from
which CO emission and absorption have been detected (Bania, Marscher, \& Barvainis 1991; Marscher, Bania, \& Wang 1991; Lucas \& Liszt 1993).  The equivalent atomic hydrogen column density of the CO
cloud is $N_H\approx 1.6\times 10^{21}$ cm$^{-2}$ (Lucas \& Liszt 1993). Thus, the total absorbing column  density in the direction of BL Lac consists of two components,  that associated with neutral hydrogen, $1.8\times 10^{21}$ cm$^{-2}$ inferred from the 21 cm measurements of Dickey et al. (1983)  and the molecular component,  yielding the total column $\sim 4.6\times 10^{21}$ cm$^{-2}$. 
For {\it Beppo}SAX data we found the $N_H$ value in a wide range (2.1 -- 7.5)$\times 10^{21}$ cm$^{-2}$  depending on an applied model and the source spectral state (see Table~\ref{tab:par_sax}), which is in agreement with  the total column value from the aforementioned{ radio} 
measurements. The residuals shown in  Figure~\ref{BeppoSAX_spectra} indicate that the observed spectra are in good agreement with the $BMC$ model. Thus,  we fitted all  observed  spectra using a $Beppo$SAX neutral column range. We note, that the $N_H$ value of $2.6\times 10^{21}$ cm$^{-2}$ is mostly suitable for the rest (long-term $RXTE$ and $Swift$ observations), which is also obtained as the best-fit  column $N_H$ for $ASCA$ observations
(see also Madejski et al, 1999; Sambruna et al., 1999). 

%
%

\begin{figure*} 
     \includegraphics[width=17.5cm]{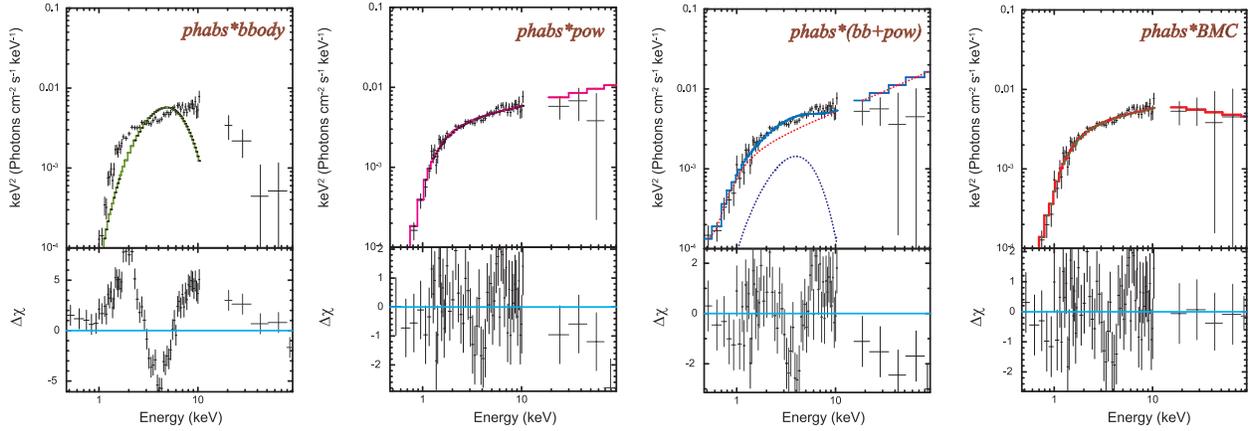}
\caption{
{
The best-fit spectra of BL~Lac observed with $Beppo$SAX during the soft state in 1999 transition (dataset ``S3'') in $E*F(E)$ 
units for the model fits (from left to right): phabs*bbody (green line, $\chi^2_{red}=1.3$ for 79 dof), 
phabs*powerlaw (purple line, $\chi^2_{red}=2.31$ for 79 dof), 
phabs*(bbody+powerlaw) (light-blue line, $\chi^2_{red}=1.78$ for 77 dof) and 
phabs*BMC (red line, $\chi^2_{red}=1.17$ for 77 dof). The data are shown by black crosses. For an additive 
model, phabs*(bbody+powerlaw), the model components are presented by dashed blue and red lines for Blackbody and  
 Powerlaw, respectively (see details in Table~\ref{tab:par_sax}). 
}
}
\label{s3_sax_dif_models}
\end{figure*}

%
%

  \begin{figure*}
 \centering
    \includegraphics[width=16cm]{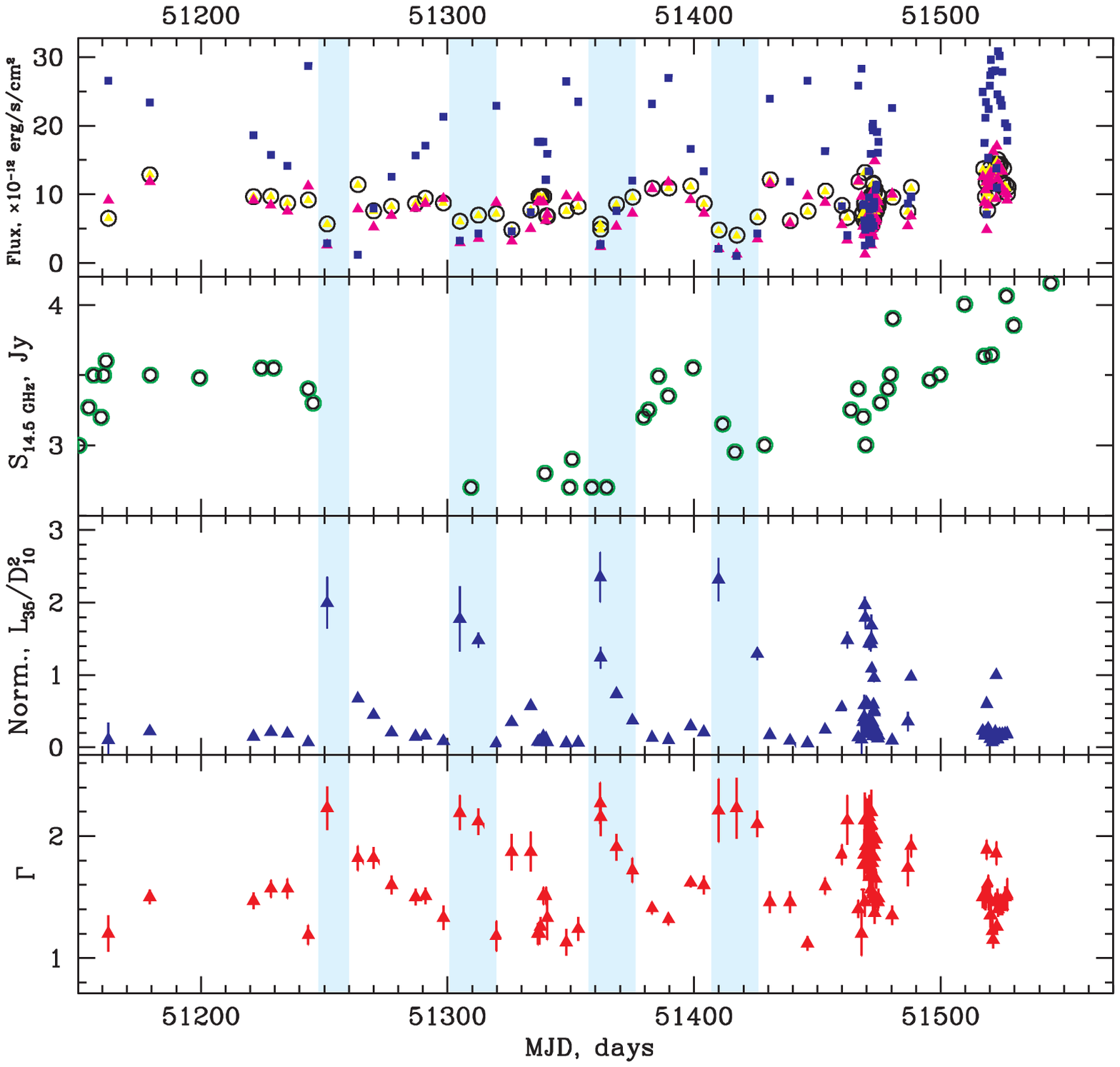}
      \caption{
{\it From Top to Bottom:}
Evolutions of the  model flux in  the 3 -- 10 keV, 10 -- 20 keV, and 20 -- 50 keV ranges (yellow,crimson, and blue 
points, respectively) using  $RXTE$/PCA, the flux density $S_{14.5 GHz}$ at 14.5 GHz (UMRAO), 
the BMC normalization and  $\Gamma$ during the
1999 flare transition ({R3, R5}). Blue vertical strips indicate the phases, when X-ray flux ($E>$ 3 keV)  anticorrelates with $\Gamma$ and the normalization $N_{BMC}$.
}
\label{lc_1999}
\end{figure*}

\subsubsection{Details of  spectral modeling \label{model choice}}

We obtained that the absorbed single power-law (phabs*power-law) model fits   well  the low and high states data only  
for ``S1'' spectrum, $\chi^2_{red}$=1.15 (80 d.o.f.), and ``S5'' spectrum, $\chi^2_{red}$=1.05 (116 d.o.f.) (see 
Table \ref {tab:list_SAX} for the notations of observations, S1-S5, and   Table~\ref{tab:par_sax} for the details of the spectral fits). We establish that the power-law model gives unacceptable fit quality, $\chi^2$  
 for all ``S2 -- S4" observed spectra  of $Beppo$SAX , in which a simple power-law model produces a hard excess. 
These significant positive residuals at high energies, greater than 10 keV, suggest the presence of additional 
emission  components in the spectrum. 
Moreover,  the thermal model (Blackbody) gives us even worse fits. 
As a result  we  tried to  check a sum of blackbody and power-law  models. In this case the model parameters  
are  $N_H=(2.7 - 4.7) \times 10^{21}$ cm$^{-2}$; $kT_{bb}=0.37-1.02$ keV and $\Gamma=1.46 - 2.4$ (see more details 
in Table~\ref{tab:par_sax}). The best fits of the {\it Beppo}SAX spectra have been found  using  the  
{Bulk Motion Comptonization model} ({BMC XSPEC} model, \cite{tl97}),   
for which $\Gamma$ ranges from 1.8 to 2.2  for all observations (see Table~\ref{tab:par_sax} 
and Figures~\ref{BeppoSAX_spectra}-\ref{s3_sax_dif_models}). 
{
Figure \ref{s3_sax_dif_models} shows the best-fit spectrum of BL~Lac observed by $Beppo$SAX during the soft state in the 1999 transition (dataset ``S3'') presented  in $E*F(E)$ units using different frame models  (from left to right): phabs*bbody 
(green line, $\chi^2_{red}=1.3$ for 79 dof), 
phabs*powerlaw (purple line, $\chi^2_{red}=2.31$ for 79 dof), 
phabs*(bbody+powerlaw) (light-blue line, $\chi^2_{red}=1.78$ for 77 dof), and 
phabs*BMC (red line, $\chi^2_{red}=1.17$ for 77 dof). The data are shown by black crosses. In particular,  for an additive 
model phabs*(bbody+powerlaw), the blackbody and   powerlaw components  are presented by blue and red  dashed lines, respectively (see details in Table~\ref{tab:par_sax}).
} 
We should emphasize that all 
$Beppo$SAX best-fit results are found using the same model (BMC) for 
the high and low states.

Our $Beppo$SAX data analysis provides strong arguments in favor of the BMC model  
to describe X-ray spectral evolution of BL Lac throughout all spectral states. Thus, we decided to analyze all available spectral data of BL Lac   using the BMC model. We provide a short description of the BMC model  in  the Introduction.
As one can see,  the BMC  has the main parameters, $\alpha$, $A$, the seed blackbody temperature $T_s$ 
and the BB normalization, which is proportional to the seed blackbody luminosity and inversely proportional to $d^2$ where 
d is the distance to the source   (see also TS16). 
{ We also apply a multiplicative $phabs$ component characterized by an equivalent hydrogen column, $N_H$ in order to take 
into account  an absorption by neutral material.   

Thus, using the same model, we carried out the spectral analysis of $ASCA$, $Suzaku$, $Swift/XRT$ and $RXTE$ observations  
and found that BL Lac was in the three spectral states (LHS, IS, HSS). The best-fit 
$\Gamma$-values  are presented in 
Tables~\ref{tab:par_swift} and \ref{tab:par_rxte}
and in Figures~\ref{BeppoSAX_spectra}-\ref{saturation}.
An evolution 
between the low state and high state is accompanied by 
a monotonic increase of the normalization parameter $N_{BMC}$ 
from 0.5 to 70$\times L_{34}/d_{10}^2$ erg/s/kpc$^2$ and by an increase of  $\Gamma$ 
from 1.1 to 2.2
(see  Figure~\ref{saturation}). 
Here, we use $L_{34}$ and $d_{10}$ as notations for soft photon luminosity in units of $10^{34}$ erg s$^{-1}$ and the distance to the source in units of 10 kpc, respectively  (see also Table \ref{tab:par_sax}). 

Note, that during the {\it RXTE} observations (from 1997  to 2001), the source was   in the IS, LHS, and
HSS, for approximately  75\%, 20\%, and 5\%  of the time, respectively.

In Figure~\ref{BeppoSAX_spectra} we demonstrate three representative $EF_E$ spectral  diagrams 
for different states of BL Lac. Data are taken from $Beppo$SAX observations 5004600400 (left panel, S1 set, LHS), 5088100100 (central panel,  S2  set, IS), and 511650011 (right panel, 
S5  set, HSS). The data are represented by black crosses and  
 the spectral model is displayed  by a red line. 
In the {bottom panels} we show  the corresponding $\Delta \chi$ versus photon energy (in keV). 

The best-fit model parameters for the HSS  (right panel, S5) are $\Gamma$=2.2$\pm$0.2,  
$N_{BMC}$=(3.08$\pm$0.06) $L_{34}/d^2_{10}$ erg/s/kpc$^2$, 
$kT_s=50\pm 10$ eV and $\log(A)$=0.35$\pm$0.07 [$\chi^2_{red}$=1.08 for 114 d.o.f], 
while the those parameters for the IS (central panel, S2) are 
$\Gamma$=2.07$\pm$0.09, $N_{BMC}$=(0.6$\pm$0.2) $L_{34}/d^2_{10}$ erg/s/kpc$^2$, 
$kT_s$=108$\pm$9 eV and $\log(A)$=0.24$\pm$0.09 
[$\chi^2_{red}$=0.87 for 80 d.o.f]; and 
those for the LHS  (left panel, S1) are $\Gamma$=1.8$\pm$0.1, $N_{BMC}$=(0.36$\pm$0.09)  
$L_{34}/d^2_{10}$ erg/s/kpc$^2$, 
$kT_s$=73$\pm$5 eV and $\log(A)$=-0.32$\pm$0.04 
[$\chi^2_{ref}$=1.06 for 78 d.o.f]. 

%
%

  \begin{figure*}
 \centering
     \includegraphics[width=11cm]{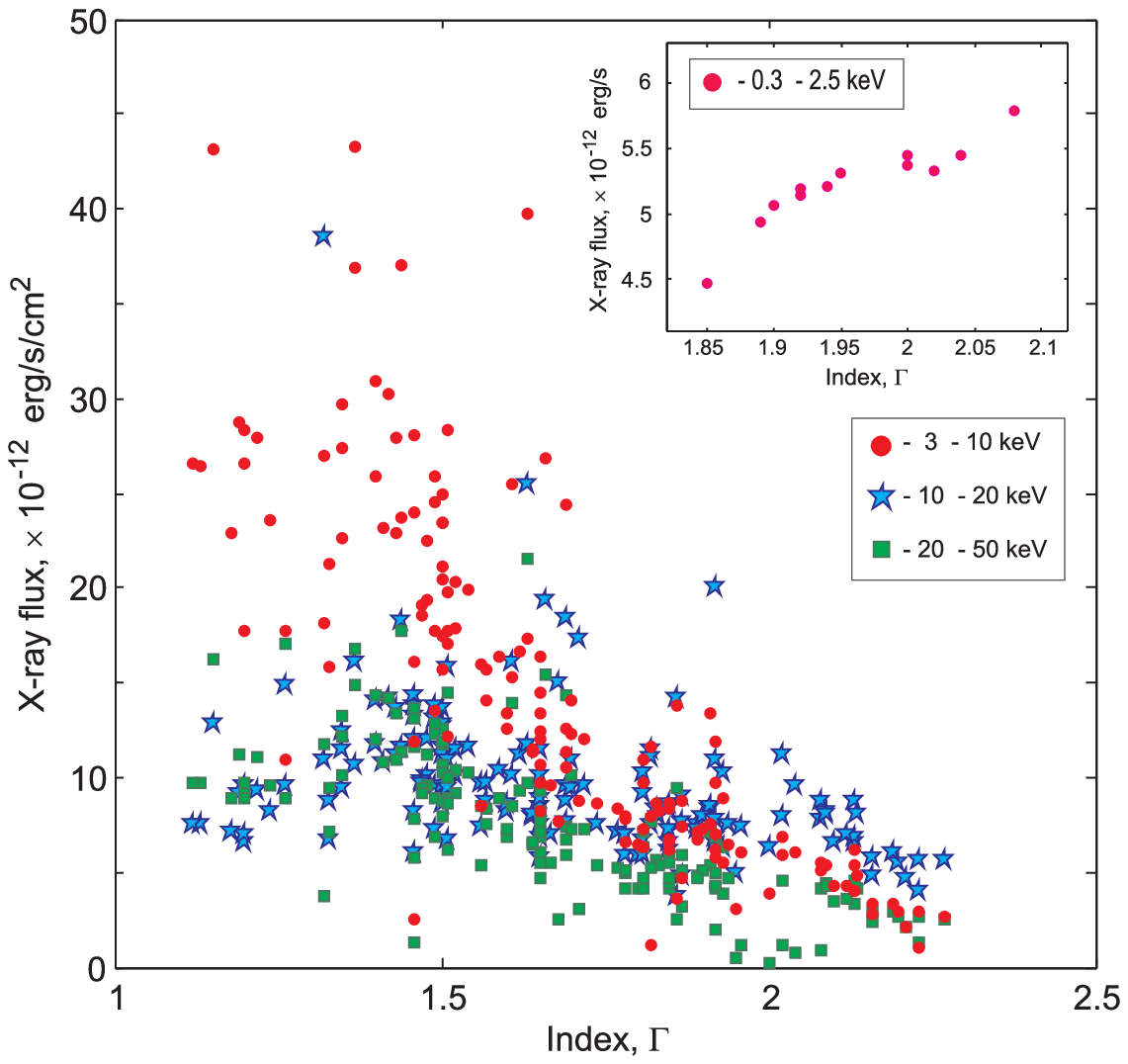}
      \caption{
Hard X-ray flux ($E>$ 3 keV) versus the photon index for the 3 -- 10 keV (red circle), 10 -- 20 keV (blue stars) and 
20 -- 50 keV (green squares) using  the {\it RXTE} observations of BL Lac (1997 -- 2001). Soft X-ray flux [0.3 -- 2.5 keV] versus the
photon index  is plotted in the incorporated panel (top right) using   $Suzaku$ and $Swift$ observations (see also Tables \ref{tab:list_SAX_Suzaku}-\ref{tab:list_Swift}).
}
\label{gamma-flux}
\end{figure*}

 \begin{figure*}
 \centering
 \includegraphics[width=6cm]{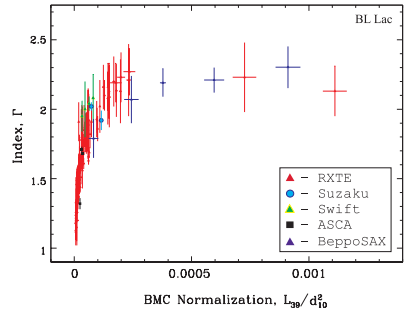}
   \caption{
Correlations of  $\Gamma$  
versus  the BMC normalization, $N_{BMC}$ (proportional to mass accretion rate) in units of $L_{39}/d^2_{10}$. 
}
\label{saturation}
\end{figure*}

Thus, we 
obtain that the seed temperatures, $kT_s$ of the $BMC$ model  vary from 50 to 110 eV. 
We also find  that the parameter $\log(A)$ of   the  $BMC$ component varies in a wide range 
between -0.86 and 0.24, (the illumination fraction $f=A/(1+A)$) and  thus,   $f$ undergoes drastic changes during an outburst phase 
for all observations. 

We should  point out the fact that 
all of the  HSS, IS and LHS spectra are characterized by a strong soft {blackbody} (BB) component  at low energies 
and a power law extending up to 100 keV, which is in good agreement with the Comptonization of soft photons  (see, e.g., Sunyaev \& Titarchuk (1980) and  TLM98) for an X-ray emission  origin. 

For the {\it Beppo}SAX observations (see Tables \ref{tab:list_SAX}, \ref{tab:par_sax}) we find that the spectral 
index $\alpha$ monotonically increases from 0.8 to 1.2 (or   $\Gamma$ from 1.8 to 2.2),  
when the normalization  of $BMC$ component (or mass accretion rate)  increases  by a factor of 8. 
We illustrate this index versus mass accretion rate correlation  in Figure~\ref{saturation} (see $blue$ triangles).

From Figure~\ref{saturation}, it is also seen that  $\Gamma,$ as a function of the normalization 
parameter $N_{BMC}$ , displays a strong saturation part at high values of the normalization $N_{BMC}$ (which is 
proportional to mass accretion rate). 
It is interesting that previous spectral analyses of $Swift$ X-ray data for BL Lac (Raiteri et al, 2013) show that
the $\Gamma$ values are scattered between 1.32 (hard spectrum) and 2.37 (soft spectrum) without correlation with the flux.

{It is worth noting that} Wehrle et al. (2016) combined{ these} $Swift$/XRT data in their low, medium, and high states in order to determine whether or not
the energy index changed when the source was brighter or fainter. They found  that 
brighter states tended to have harder X-ray spectra. 
They revealed the same result  based on $RXTE$ 
data in the 2 -- 10 keV range. 
Specifically, they argued that the spectral index tended to flatten when the source was bright, and conversely, became steeper when the source was faint.  

However, we should  notice that all{ Wehrle's} 
spectral studies used a simple power-law model and the brightness was related to the 2 -- 10 keV range. 
Therefore, the power-law index   was inspected as a function of relatively hard 
X-ray brightness (Wehrle et al., 2016). In contrast, in Figure~\ref{gamma-flux} we tested the behavior 
of  $\Gamma$  in the framework of the $BMC$ model and investigated  the $\Gamma-Flux$ 
evolution for different energy bands: the 3 -- 10 keV ($red$ circle), 10 -- 20 keV ($blue$ stars) and 
 20 -- 50 keV ($green$ squares) using $RXTE$ observations BL Lac (1997 -- 2001), as well as the 0.3 -- 2.5 keV  
($pink$ circles; see incorporated top right} panel) using $Swift$ and $Suzaku$ observations. From this 
Figure  one can clearly see  that  hard X-ray flux ($E>$ 3 keV) decreases with respect to $\Gamma$  (see also 
Wehrle et al., 2016), while  softer 
X-ray flux ($E<2.5$ keV)  increases when   $\Gamma$  increases (see the incorporated panel). Thus, we should emphasize that the $\Gamma-Flux$  relation for BL Lac is an energy dependent correlation.

To make  {the \it RXTE} data analysis we used information obtained
using  the $Beppo$SAX, $ASCA$ and $Suzaku$ best-fit spectra, which can provide well calibrated spectra at soft energies ($E<$~3 keV). 
Because the {\it RXTE}/PCA detectors cover energies  
above 3 keV, for our analysis of the {\it RXTE} spectra we fixed a key parameter of the {\it BMC} model  
($kT_{s}=$ 70 eV) obtained as a mean value  of 
$kT_s$ in  our   analysis of the {\it Beppo}SAX spectra.

 In Figure~\ref{lc_1999}, from the top to the bottom we demonstrate
 evolutions of the model flux in the 3 -- 10 keV, 10 -- 20 keV, and 20 -- 50 keV energy ranges (yellow, crimson, and blue 
points, respectively), the 
flux density $S_{14.5 GHz}$ at the 14.5 GHz (UMRAO, Villata et al., 2009), the BMC normalization, and $\Gamma$  for the 1999 flare transition  ({R2}). Blue vertical strips mark intervals 
when hard X-ray emission ($E>3$ keV) 
anticorrelates with  $\Gamma$ and the normalization $N_{BMC}$.

 The spectral evolution of BL Lac was previously investigated using {some of} the $Swift$ 
 data (see for example, $Sw1$, $Sw3$ and $Sw4$ in {Table~\ref{tab:list_Swift}}) 
by many authors. In particular, Wehrle et al. (2016) analyzed the 2012 -- 2013 
{$Swift$ data (partially, $Sw1$ and $Sw3$) and the long (2005 -- 2011) $RXTE$ observations,
while Raiteri et al. (2010) studied the $Swift$ observation of BL~Lac during the 2008 -- 2009  period ($Sw1$, $Sw3$ and $Sw4$). Furthermore, 
Raiteri et al. (2013) reexamined the $Swift$ observations of BL~Lac during the period  2008 -- 2012 
and compared them with the $RXTE$ observations of BL~Lac  for  the same  period.  

 Wehrle et al. (2016) modeled their $Swift$ data using a single power-law continuum and 
Galactic absorption for which  hydrogen column density $N_H = 3.4\times 10^{21}$ cm$^{-2}$. They described the 
spectrum of BL~Lac for the low, medium, and high source states with different spectral indexes, $\alpha_{X}\sim 0.97$, 
0.86, and 0.67 ($\alpha_X=\Gamma_X-1$), respectively. They concluded that the spectral index  decreased when 
the source  made  a transition from low to high states. Then, Wehrle et al. (2016) checked this behavior  using the $Swift$ data sets in combination with $NuSTAR$ (3 -- 70 keV) observations of BL~Lac, 
applying three models: power law [phabs*pow], broken power law  [phabs*bknpow], 
and log-parabolic [phabs*logpar] models.  
As a  result, assuming a power-law model with  fixed $N_H$, they found acceptable fits with $\Gamma\sim 1.9$, on average.
The broken power-law and log-parabolic models improved 
the fit quality, with $\Gamma\sim 2$, on average. It is interesting that  results of these
fittings  suggested that the observed X-ray spectrum of BL~Lac might be steeper at softer photon energies than at harder  ones. 

Raiteri et al. (2010) tested  the $Swift$ data set (2008 -- 2009, partly $Sw1$, $Sw3$ and $Sw4$) 
using their spectral analysis.  First they  fitted spectra using a single power law
with free absorption, and then they fixed the Galactic absorption at $N_H=3.4\times 10^{21}$ cm$^{-2}$. Due to poor statistics, they applied  a double power law model to BL~Lac spectra. 
Although, fits using free absorption gave  very variable $N_H$, which corresponded
 to unreal changes of absorption. Therefore, they favored the second model 
with fixed $N_H$ and  acceptable  $\chi^2$. For some spectra, a double power law model with 
absorption fixed to the Galactic value definitely improved the fit quality.  Raiteri et al. (2010) 
argued that  $\Gamma$ ranges from 1.9 to 2.3, indicating that these spectra varied from  hard to soft  (with the average value $\Gamma\sim2$).
Finally, Raiteri et al. (2013) refitted the BL Lac spectra  (around $Sw1$, $Sw3$, $Sw4$ periods) applying an absorbed  power law with the same value of $N_H$ as that used by Wehrle et al. As a result, they found that  the source made a transition  from the hard state ($\Gamma$=1.3) to the soft state ($\Gamma$=2.4) without  any correlation with the source flux.
As one can see, Wehrle's and Raiteri's studies  did not 
account for the low-energy excess, particularly in the soft state of the source.

}

}
We have also found a similar 
spectral behavior using our BMC model along with the full set of the $Swift$ observations.  
In particular, as in the aforementioned Wehrle's and Raiteri's et al. papers, we  also reveal that 
BL Lac demonstrates the quasi-constancy of $\Gamma$ during the IS -- HSS transition.
Furthermore, we find that  $\Gamma$  strongly  saturates  at  2.2 
at high values of $N_{bmc}$ 
(or at high values of  the mass accretion rate). 

In the LHS, the seed photons  with lower $kT_{s}$, 
related to lower mass accretion rate,  
are Comptonized more efficiently because the illumination fraction $f$ (or  $\log(A)$) is higher.
In contrast, 
in the HSS, these parameters, $kT_{s}$ and $\log(A)$  show an opposing behavior; namely 
 $\log(A)$ is lower for higher $kT_s$.  This means that 
a relatively small fraction of the seed photons, whose temperature is higher because of the higher mass accretion rate in the HSS than in the LHS, is  Comptonized.

Thus, our spectral model shows  very good performance through
all data sets. 
{In Tables~\ref{tab:par_swift} and \ref{tab:par_rxte} 
we demonstrate a good performance of the BMC model in application to the $ASCA$, $Suzaku$, $Swift$ and $RXTE$ data.}
The reduced  $\chi^2_{red}=\chi^2/N_{dof}$ 
(where $N_{dof}$ is the number of degrees of freedom) is  less; approximately  1 ($0.72<\chi^2_{red}<1.15$) 
for  all 
observations. 

}
\section{Discussion \label{disc}}

Before proceeding with  an interpretation of the observations,
let us briefly summarize them as follows. 
i) The spectral data of BL Lac   are well fitted by the BMC model for all 
analyzed LHS and HSS spectra (see e.g., Figure~\ref{BeppoSAX_spectra} and
Tables~\ref{tab:par_sax}, \ref{tab:par_swift} and \ref{tab:par_rxte}).
ii) The Green's function index of the BMC component $\alpha$ (or 
$\Gamma$)  monotonically rises and saturates with an increase of the BMC normalization
(proportional to $\dot M$). The photon 
index saturation level of the BMC component is approximately 2.2 (see
Figure~\ref{saturation}). 
iii) Blazar BL Lac  undergoes spectral state transitions during X-ray outburst events. 
The X-ray evolution of BL Lac is characterized by a number of
similarities   with respect to those in  Galactic BHs (GBHs). 
For example, the X-ray spectral index of BL~Lac demonstrates a
saturation phase in its soft state in the same manner as that in  GBHs. In the soft states of  GBHs we do not see their jets and outflow (associated with radio emission, see, e.g., Migliari \&
Fender 2006).

Below, in Sect.~4.1, we demonstrate some episodes in which radio emission is
completely suppressed during  the X-ray soft state.  
Quenching of the radio emission in the soft state of BL Lac is in agreement 
with that  found in GBHs (see  e.g., GRS~1915+105 (TS09)).
Furthermore,  BL Lac is well known by its prominent variability in optical (see Larionov et
al., 2010; Gaur et al., 2015) and radio bands (Wehrle et al., 2016). Thus, the correlations between
optical and  radio  emissions are deeply investigated. 
However, the study of the BL Lac variability in X-rays and its possible correlations  (or anti-correlations) between  X-ray and radio emissions have  only recently been developed . While the
X-ray has a relatively narrow  energy range (0.3 -- 150 keV) in comparison with the broad-band SEDs, its variability points to the broad-band variability of BL Lac.   The X-ray part of the spectrum  is intermediate between two global peaks: at low energies (optical/IR--UV)  
and high-energies (up to $\gamma$-rays).
Below we  investigate the connection
between the radio flux density  and X-ray flux and  we find that it is qualitatively similar to that found in GBHs.

%
%

\subsection{Connection between radio and X-ray emission in BL Lac}

The 230 GHz (1.3 mm) light curve was obtained using the Submillimeter Array (SMA) in Mauna Kea (Hawaii), see
Villata et al. (2009).
 In Figure~\ref{lc_1999} 
we present an evolution of the flux density $S_{14.5 GHz}$ at 14.5 GHz (UMRAO\footnote{University of Michigan Dadio Astronomy Observatory Data
Base, http://www.astro.lsa.umich.edu/obs/radiotel/umrao.php.}, Villata et al., 2009, Aller
et al. 1985)
 along with  a spectral parameter evolution such as the BMC normalization and 
$\Gamma$ during the 1999 flare transition set ({R2}). Blue vertical strips indicate the  phases 
when the radio flux ($S_{14.5 GHz}$) anticorrelates with  $\Gamma$ and the normalization, $N_{BMC}$. 
We should point out a clear anti-correlation between  the radio flux density and  $\Gamma$,
similar to that identified in some Galactic microquasars, for example  in GRS~1915+105  (see discussion below). 

Due to lack of the soft X-ray monitoring, $E<3$ keV. During the {\it RXTE} observations 
we associate soft  X-ray evolution with the evolution of $N_{BMC}$ (which is proportional to the seed soft photon flux).
We  formulate the most important  results for BL Lac: (i) The strong radio flare occurs usually on the eve 
of  X-ray flare and (ii) during proper X-ray flare  the radio flux density  significantly decreases. 
For example, for the {\it  RXTE} data (1999, set R3, R5), the MJD intervals of strong $N_{BMC}$ (soft X-ray flare) coincide with 
low radio flux density $S_{14.5 GHz}$ ,  indicated by blue vertical strips (e.g., centered on MJD 51250, 51305, 
51350, 51415, 51475,  see Fig.~\ref{lc_1999}). It is interesting that these intervals are accompanied by the increase 
of $\Gamma$   above $2$. 
{A similar relation between radio (230 GHz) and hard X-ray behavior in BL~Lac  
was also found for the $Swift$ observations from October to- November 2012. 
}

In Figure~\ref{radio}, from the top to the bottom we show 
 evolutions of the 230 GHz/345 GHz (pink/green flux densities (see also Wehrle et al., 2016)); flux in the 1.5 -- 10 keV 
(black points); and flux in the 0.3 -- 1.5 keV (red 
points) 
as a function of   MJD time. 
{
From this Figure   one can see that the  X-ray flare (1.5 -- 10 keV, see MJD 56229) is developed at the low level of radio flux density, which can also indicate a possible episode of anticorrelation between the radio and X-ray 
emissions observed in BL~Lac. 
}

This anticorrelation between the radio and X-ray emissions observed in BL Lac 
 suggests a inflow/outflow  scenario.
Strong radio flux is usually associated with a powerful outflow in the form of jet or outflow (wind). In  contrast,  the X-ray 
emission is  accompanied by a powerful accretion inflow.  
 Thus, the outflow and inflow  effects lead to anticorrelation 
between corresponding  X-ray and radio emissions.  During the inflow episode (converging inflow case), an accretion 
is seen  as  X-ray emission
while radio emission is suppressed. Alternatively, for the outflow event 
(a divergent flow case), the radio emission is dominant and thus, inflow (accretion) observed in X-rays
is suppressed. 
Therefore, divergent and converging cases of the flow around the central object  relate to the  corresponding regimes of the radio and X-ray dominances.

{ Radio  emission (at 230 GHz and 345 GHz) is more variable  than   emission in the X-ray range observed by $Swift$ during the period October -- November 2012 (see Fig.~\ref{radio}).
 This  might suggest  possible different origins of radio and soft X-rays.
Radio emission could be related to  jet  blobs (knots), while  soft X-rays emerge  from the inner part of accretion disk. 
}

\subsection{Saturation of the  index as a signature of a BH  \label{constancy}}

%
%
\begin{figure}
\begin{center}
   \resizebox{\hsize}{!}{\includegraphics{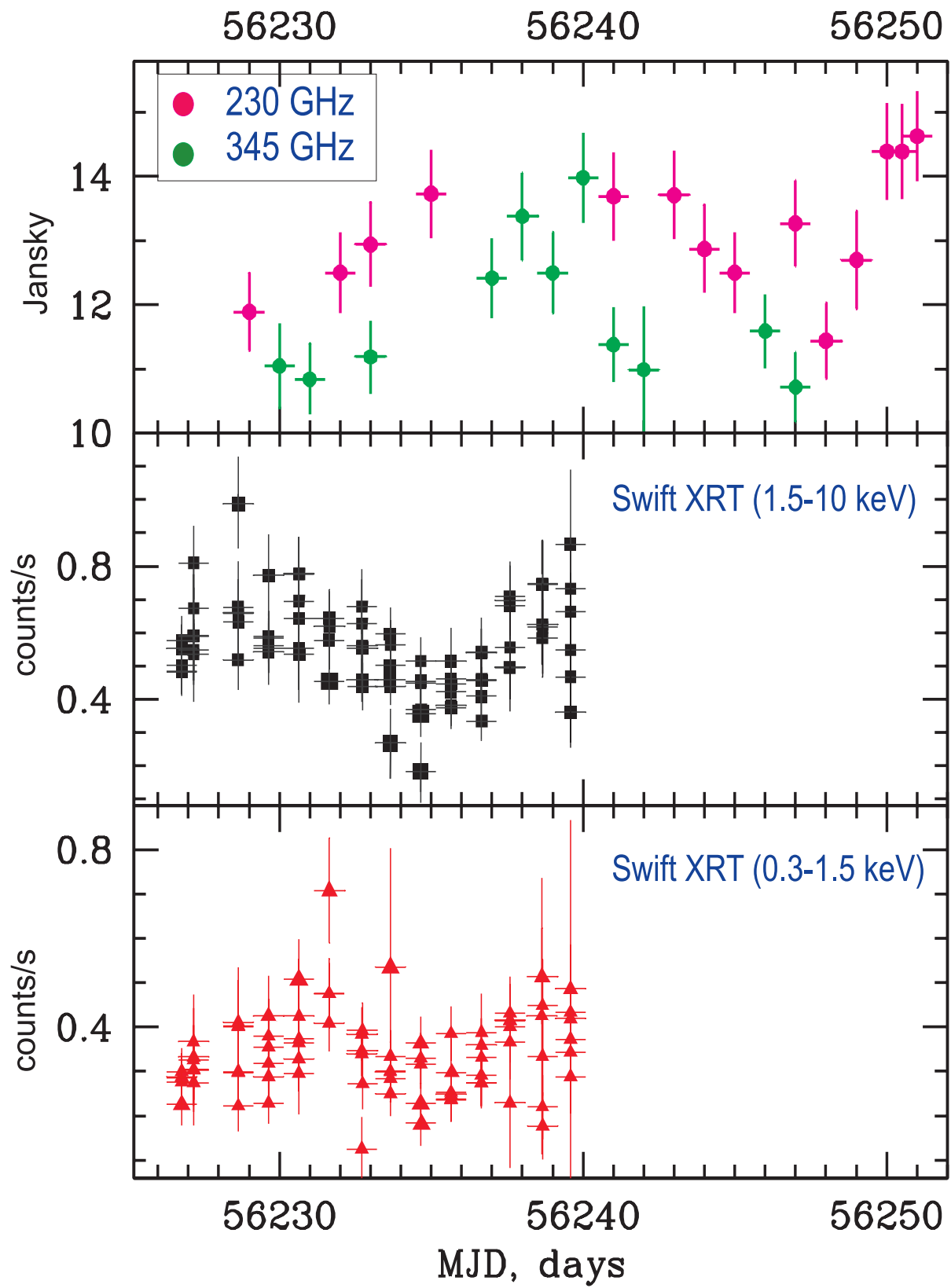}}
\end{center}
  \caption{
{\it From Top to Bottom:}
Evolution of the flux density $S_{230 GHz}$ (pink) and $S_{345 GHz}$ (green points)  at 230 GHz and 345 GHz (SMA), 
the model flux in the 1.5 -- 10 keV (black points, Swift XRT), the model flux in the 0.3 -- 1.5 keV (blue points, Swift XRT), 
and X-ray hardness ratio HR. 
}
\label{radio}
\end{figure}

We establish that  $\Gamma$ correlates with the  BMC normalization,  $N_{BMC}$ (which is proportional to $\dot M$)
and finally saturates at high values of $\dot M$ (see Figure~\ref{saturation}). 
Titarchuk \& Zannias (1998) developed the semi-analytical theory of X-ray spectral formation in the converging flow into a  BH. They demonstrated that the spectral index of the emergent X-ray spectrum saturated at high values of mass accretion rate (at higher than the Eddingtion one). Later analyzing  the data of  {\it RXTE}  for many black hole candidates (BHs)   ST09  and  \cite{tsei09}, \cite{ST10} and STS14 demonstrated that this  index saturation effect was seen  in many  Galactic BHs (see for example, GRO J1655-40, GX 339-4, H1743-322, 4U 1543-47, Cyg X-1,  XTEJ1550-564, GRS 1915+105 ).  The levels of the index saturation are  at different values which presumably depend on the plasma temperature of the converging flow [see Monte Carlo simulations by Laurent \& Titarchuk (1999), (2011)].  
 
For our particular source, BL Lac,  we also reveal  that   $\Gamma$ 
monotonically increases  from 1.2  and then finally saturates at a value of 2.2  (see  Figure~\ref{saturation}).  
Using the  index-$\dot M$ correlation found in BL Lac  we can  estimate a BH mass  in 
this source  using scaling of this correlation with those   detected  
in a  number of GBHs 
(below see the details). 


\subsection{An estimate of BH mass in BL Lacertae \label{bh_mass}}

To estimate the BH mass, $M_{BH}$ , of BL Lac, we chose two galactic sources, 4U~1543--47 and GX~339--4 
(see ST09), as the reference sources
whose BH masses {($M_{1543}=9.4\pm 1.0$ M$_{\odot}$, see Orosz 2003; $M_{339}>6$ M$_{\odot}$,  see Mu$\tilde n$oz-Dariaz et al. 2004)} 
and  distances {($d_{1543}=9.1\pm 1.1$ kpc, see Orosz et al., 1998; $d_{339}=7.5\pm 1.6$ kpc, see Hynes et al., 2004)} 
have now been well established  (see Table~\ref{tab:par_scal}).   
The  BH mass  in  4U~1543--47 was  also estimated applying dynamical methods { (Orosz, 2003)}. 
For  a BH mass estimate of BL Lac we  used the BMC normalizations, $N_{BMC}$ of these reference sources.  

Thus, we scaled  the index versus  $N_{BMC}$  correlations for these reference sources  with that of 
the target source  BL Lac (see Fig.~\ref{three_scal}). 
The value of the  index saturation  is   almost the same, $\Gamma\sim 2.2,$  for all these target and reference sources.  
We applied the correlations found in  these two reference sources to make  a  
 comprehensive cross-check of  a BH mass estimate for BL Lac.

%
%
\begin{figure*}
\begin{center}
 \includegraphics[width=14cm]{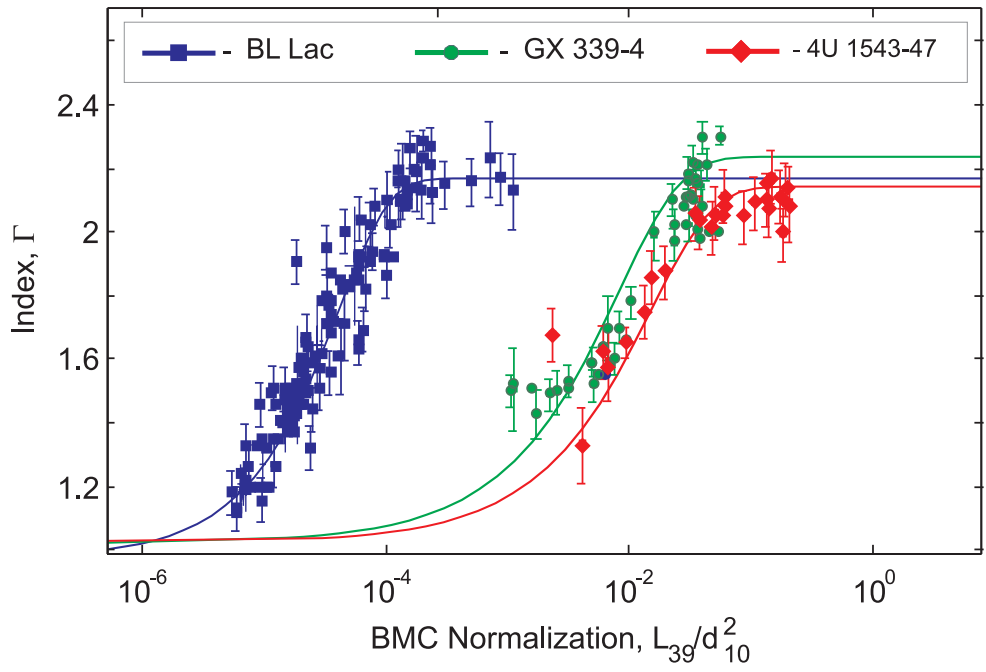} 
\end{center}
  \caption{
Scaling of  $\Gamma$ versus the normalization $N_{BMC}$ for BL Lac (blue squares 
-- target source) using those correlations for the 
Galactic
reference sources,  GX~339--4 
(green circles) and 4U~1543--47 (red diamonds). 
}
\label{three_scal}
\end{figure*}

As one can see from Figure \ref{three_scal}, the correlations of the target source (BL Lac) and the reference 
sources have  similar shapes and index saturation levels.
Hence, 
it allows us to make 
a reliable scaling of these correlations with that of  BL Lac.
The scaling procedure was implemented in a similar way as in ST09, Titarchuk \& Seifina 2016a, 2016b, hereafter TS16a and TS16b.   
We introduce an analytical 
approximation  
of the $\Gamma(N_{bmc})$ correlation, 
fitted by a function 

\begin{equation}
{\cal F}(x)= {\cal A} - ({\cal D}\cdot {\cal B})\ln\{\exp[(1.0 - (x/x_{tr})^{\beta})/{\cal D}] + 1\}
\label{scaling function}
\end{equation}
with $x=N_{bmc}$.

Fitting  of the  observed correlation  by  this function ${\cal F}(x)$
provides us a set of the best-fit parameters $\cal A$, $\cal B$, $\cal D$, $x_{tr}$, and $\beta$.  
A more detailed description of these parameters is given in TS16a.

In order to implement 
this BH mass determination for the target source one should rely on 
the same shape of the $\Gamma-N_{bmc}$ correlations for the target source and those for the reference sources.  To estimate BH mass,  $M_t$  , of BL Lac (target source), one should slide the reference source correlation 
along the $N_{bmc}-$axis  to that of the target source (see Fig. \ref{three_scal}),

\begin{equation}
M_t=M_r \frac{N_t}{N_r}
\left(\frac{d_t}{d_r}
\right)^2 f_G,
\label{scaling coefficient}
\end{equation}
\noindent where t and r correspond to the target and reference sources, respectively
and a  geometric factor,  
$f_G=(\cos\theta)_r/(\cos\theta)_t$; the inclination angles $\theta_r$,  
$\theta_t$ and $d_r$, $d_t$ are distances to the reference and target sources, respectively (see ST09). 
One can see values of $\theta$ in  
Table \ref{tab:par_scal} and  if some of these $\theta$-values  are unavailable then we assume that 
$f_G\sim1$.

In Figure~\ref{three_scal} we demonstrate   the $\Gamma-N_{bmc}$ correlation  for BL Lac 
(blue squares) 
obtained  using  the $RXTE$, $ASCA$, $Suzaku$, $Swift$ and $Beppo$SAX spectra along with the correlations for the two 
Galactic reference sources, GX~339--4 (green circles) and 4U~1543--47 (red diamonds). BH masses and distances for each 
of these target-reference pairs are presented  in Table~\ref{tab:par_scal}. 

A BH mass, $M_t$  , for BL Lac can be evaluated using  the formula (see TS16a)
\begin{equation}
M_t= C_0 {N_t} {d_t}^2 f_G 
\label{C0 coefficient}
,\end{equation}
\noindent where 
$C_0=(1/d_r^2)(M_r/N_r)$ is the scaling coefficient for each of  the pairs (target and reference sources), 
masses $M_t$ and $M_r$ are in solar units, and $d_r$ is the distance to a particular reference source  measured in kpc.

We use values of $M_r$, $d_r$, $d_t$, and $\cos (i)$ from Table~\ref{tab:par_scal} 
and then  we calculate the lowest limit of the mass, using the best fit value of  $N_t= (9.6\pm 0.1)\times 10^{-5}$ 
taking them at the beginning of the index saturation  (see Fig. \ref{three_scal}) and measuring
in units of $L_{39}/d^2_{10}$ erg s$^{-1}$ kpc$^{-2}$ [see Table \ref{tab:parametrization_scal}
 for values of the parameters of function ${\cal F}(N_t)$ (see Eq. 1)].
Using $d_r$, $M_r$, $N_r$ (see ST09) we found that  $C_0\sim 3.5$ and  $3.34$ 
for GX~339--4 and 4U~1543--47, respectively. 
Finally,  we obtain that $M_{bl}\ge 2.7\times 10^7~M_{\odot}$ 
($M_{bl}=M_t$) 
assuming $d_{bl}\sim$300 Mpc~
and  $f_G\sim1$. To determine the distance to BL Lac (2200+420) we use the formula
\begin{equation}
d_{bl}=z_{bl}c/H_0
\label{bllac_distance}
,\end{equation} 
where  the redshift $z_{bl}=0.069$  for BL Lac (2200+420) (see { Wright 2006}), 
$c$ is the speed of light and $H_0=70.8\pm 1.6$
km s$^{-1}$ Mpc$^{-1}$  is the Hubble constant.  
We summarize all these results in Table~\ref{tab:par_scal}.

It is worth noting that  the inclination of BL Lac may be different from those  for the reference Galactic sources (e.g., $i\sim 20^{\circ}$ for 4U~1543--47), therefore we take this  BH mass estimate for BL Lac as the 
lowest BH mass value  because $M_{bl}$ is a reciprocal function of $\cos (i_{bl})$
(see Eq.~\ref{C0 coefficient} taking into account that $f_G=(\cos\theta)_r/(\cos\theta)_t$ there). 

The obtained  BH mass estimate is in agreement with a ``fundamental plane'' estimate 
($M_{bl}\sim 1.7\times 10^8 M_{\odot}$, Urry et al., 2000). However,   using a minimum timescales and  
variability method Liang \& Liu (2003)  obtained a lower estimate of a BH mass value,  $M_{bl}\sim 3.1\times 10^6 M_{\odot}$. 

{
 Our scaling method was effectively applied  to find BH masses of Galactic 
(e.g., ST09, STS13) and extragalactic black holes (TS16a,b; Sobolewska \&
Papadakis 2009; Giacche et al. 2014). Recently the scaling method was successfully implemented  to estimate BH masses of  two ultraluminous X-ray (ULX) sources M101 ULX--1 (TS16a) and ESO~243--49 HLX--1 (TS16b). These findings suggest BH masses of  approximately $10^4$ solar masses in these unique objects. 

In fact,  there are a few scenarios proposed 
 for interpretation of ULX phenomena. 
First, these sources could be stellar-mass black holes (BHs), which are significantly 
less than 100 M$_{\odot}$, radiating at Eddington or super-Eddington rates (Titarchuk et al. 1997; Mukai et al. 2005). 
Alternatively, they could be intermediate-mass black holes (IMBH; more than 100 M$_{\odot}$) where the luminosity  is essentially sub-Eddington. Recently, Bachetti et al.
(2014) discussed a new scenario for ULX, in which some ULX sources can be powered by a neutron star.  
Thus, the exact origin of these objects remains uncertain and there is still no general consensus on what triggers the aforementioned ultraluminous regime. However, the mass evaluation of central sources by  the scaling method, now applied to extragalactic sources, can shed light on this problem. Furthermore, 
the scaling technique may prove to be useful for mass evaluation of other extragalactic sources 
with prominent activity, such as  active galactic nuclei and tidal disruption event sources, and so on. 
}

\section{Conclusions \label{summary}} 

We found the low$-$high state transitions observed in  BL Lac using the full set of $Beppo$SAX, $ASCA$, $Suzaku$, $RXTE$ and 
$Swift$ observations. 
We demonstrate a validity of fits of the observed spectra using the BMC model   
for all observations, independently of the spectral state of the source. 

We investigated the X-ray outburst properties of BL Lac and confirm the presence of the spectral state transition 
during the outbursts using hardness-intensity diagrams 
and the index$-$normalization (or $\dot M$) correlation observed in BL Lac, which are similar to those in Galactic BHs. 
In particular, we find that BL Lacertae follows the $\Gamma-\dot M$ correlation previously obtained for 
the Galactic BHs,  GX~339--4 and 4U~1534--47, 
 taking into account the 
particular values of the $M_{BH}/d^2$ ratio (see Fig.~\ref{three_scal}).
The photon index of the BL Lac spectrum is  in the range $\Gamma = 1.2 - 2.2$. 

 We applied the observed index-mass accretion rate correlation to estimate $M_{BH}$  in BL Lac. 
This scaling method was  successfully implemented  to find  BH masses of Galactic (e.g., ST09, STS13) 
and extragalactic black holes [TS16a,b; \citet{sp09}; \cite{ggt14}].  
We find  values of 
{
$M_{BH}\geq 3\times 10^7 M_{\odot}$.   
}  
Furthermore, our  BH mass estimate is 
in an agreement with the previous BL Lac  BH mass  estimates of $(0.3 - 17) \times 10^7$ M$_{\odot}$ evaluated using 
alternative methods  (Woo \& Urry  2002; Liang \& Liu 2003; Ryle 2008).
Combining all these estimates with  the inferred  low temperatures of the seed (disk) photons $kT_s$ 
we argue  that 
the compact object of  BL Lac is likely to be a supemassive black hole of at least 
$M_{BH}> 3\times10^7 M_{\odot}$. 

\begin{acknowledgements}

This research was performed using  data supplied by the UK $Swift$ Science Data Centre at the University of Leicester. We  acknowledge Valentina  Konnikova for her help with obtaining radio data needed for our research.  We also thank Alexandre Chekhtman  for useful scientific discussions and comments.  We  appreciate  the thorough analysis of 
the  paper  by the referee.
\end{acknowledgements}




\begin{deluxetable}{l c c c c c c}
\tablewidth{0in}
\tabletypesize{\scriptsize}
    \tablecaption{The list of the {\it RXTE} observations of BL Lac}
    \renewcommand{\arraystretch}{1.2}
%
 \label{tab:list_RXTE}
\tablehead
{Number of set  & RXTE Proposal ID &  Start time (UT) && End time (UT)  && MJD interval}
\startdata
R1 ................ & 10359, 20423    & 1997 July 16 && 1997 July 21           && 50645-50650$^{1, 2, 3}$  \\%
R2 ................ & 30213, 30255, 40171, 40174  & 1998 Dec 15 && 1999 Dec 14 && 51162-51526           \\
R3 ................ & 50181           & 2000 Apr 19  && 2001 Jan 15            && 51653-51924           \\
      \enddata
\\References: 
(1) Madejski et al. (1999); 
(2) Grove et al. (1997);  
(3) Tanihata et al. (2000). 
\end{deluxetable}


\begin{deluxetable}{l l l l l c c}
\tablewidth{0in}
\tabletypesize{\scriptsize}
    \tablecaption{The list of the {\it Beppo}SAX observations of BL Lac }
    \renewcommand{\arraystretch}{1.2}
%
 \label{tab:list_SAX}      
\tablehead{
Number of set & Obs. ID& Start time (UT)  &  End time (UT) & MJD interval & Mean count rate \\
              &        &                  &                &              &  (cts/s)     }
\startdata
S1 ................ & 5004600400 & 1997 Nov 8  00:28:13  &  1997 Nov 8  14:33:18  &50760.0 -- 50760.6$^{1,2}$ & 0.145$\pm$0.003 \\
S2 ................ & 5088100100 & 1999 June 5 08:05:40  &  1999 June 7 12:59:08  &51334.3 -- 51336.8$^1$     & 0.096$\pm$0.001 \\
S3 ................ & 5088100200 & 1999 Dec 5  11:06:31  &  1999 Dec  6 17:43:08  &51517.4 -- 51517.3$^1$     & 0.150$\pm$0.002 \\
S4 ................ & 5116500100 & 2000 July 26 10:12:39 &  2000 July 27 06:43:33 &51751.4 -- 51752.2$^1$     & 0.078$\pm$0.002 \\
S5 ................ & 5116500110 & 2000 Oct 31 20:46:55  &  2000 Nov 2 09:59:28   &51848.9 -- 51850.4$^1$     & 0.361$\pm$0.003 \\
      \enddata
\\References: 
(1) Ravasio et al. (2001); 
(2) Padovani et al. (2001). 
\end{deluxetable}


\begin{deluxetable}{l l l l l c c} 
\tablewidth{0in}
\tabletypesize{\scriptsize}
    \tablecaption{The list of the {\it ASCA} 
observations of BL Lac in the energy range 0.3 -- 10 keV used in our analysis.}
    \renewcommand{\arraystretch}{1.2}
%
 \label{tab:list_asca}      
\tablehead{Number of set & Obs. ID& Start time (UT)  &  End time (UT) & MJD interval & Mean count rate \\
              &        &                  &                &              &  (cts/s)      }
\startdata
A1 ................ & 73088000       & 1995 Nov 22  01:03:33  & 1995 Nov  22 19:51:21 & 50043.05 -- 50043.8$^1$ & 0.254$\pm$0.003 \\
A2 ................ & 15505000       & 1997 July 18 14:02:10  & 1997 July 19 14:55:29 & 50647.59 -- 50648.7$^{1, 2}$ & 0.526$\pm$0.003 \\
A3 ................ & 77000000       & 1999 June 28 04:22:50  & 1999 June 30 18:17:19 & 51357.22 -- 51359.8     & 1.787$\pm$0.002 \\
      \enddata
\\References: 
(1) Sambruna et al. (1999); 
(2) Tanihata et al. (2000).
\end{deluxetable}


\begin{deluxetable}{l l l l l c c} 
\tablewidth{0in}
\tabletypesize{\scriptsize}
    \tablecaption{The list of the {\it Suzaku} 
observations of BL Lac in the energy range of 0.3 -- 10 keV used in our analysis.}
    \renewcommand{\arraystretch}{1.2}
 \label{tab:list_SAX_Suzaku}      
\tablehead{Number of set & Obs. ID& Start time (UT)  &  End time (UT) & MJD interval & Mean count rate \\
              &        &                  &                &              &  (cts/s)      }
\startdata
Sz1 ................ & 701073010      & 2006 May 27 05:31:44  & 2006 May 28 07:20:19 & 53882.2 -- 53883.3 & 0.554$\pm$0.004\\
Sz2 ................ & 707044010      & 2013 May 19 14:51:56  & 2013 May 19 20:56:23 & 56431.6 -- 56431.9 & 0.979$\pm$0.009\\
      \enddata
\end{deluxetable}



\begin{deluxetable}{l l l l l l c}
\tablewidth{0in}
\tabletypesize{\scriptsize}
    \tablecaption{The list of $Swift$ observations of BL Lac. }
    \renewcommand{\arraystretch}{1.2}
 \label{tab:list_Swift}      
\tablehead{Number of set & Obs. ID& Start time (UT)  &  End time (UT) & MJD interval}
\startdata
Sw1 ................ & 00030720(001-031, 033-118, 120-162,  & 2006 May 28     & 2016 Nov 2   & 53883.3 -- 57694.1 \\
                     & 164, 166-189, 191-212)$^{1,2,3}$                       &              &                    \\
Sw2 ................ & 00034748(001-003)                    & 2016 Oct 6      & 2016 Oct 8   & 57667.2 -- 57669.2 \\
Sw3 ................ & 00035028(001-009, 011-014, 017, 019, & 2005 July 26    & 2014 Dec 21  & 53577.0 -- 57012.1 \\
                     & 021-028, 030-043, 045-060, 062-088,  &                 &              &                    \\
                     & 090-109, 111-146, 148-160, 162-172,  &                 &              &                    \\
                     & 174-215, 217-223, 225-255)$^{1,2,3}$ &                 &              &                    \\
Sw4 ................ &  00090042(001-022)$^{1,2}$           & 2008 Aug 20     & 2008 Sept 25 & 54698.6 -- 54734.9 \\
Sw5 ................ &  00092198(001-020)                   & 2015 May 14     & 2015 Dec 15  & 57156.1 -- 57371.6 \\
      \enddata
\\References: 
(1) Raiteri et al. (2013); 
(2) Raiteri et al. (2010); 
(3) Wehrle et al. (2016).
\end{deluxetable}

%
%

\begin{deluxetable}{lllllll} 
\tablewidth{0in}
\tabletypesize{\scriptsize}
    \tablecaption{Best-fit parameters  of the combined $Beppo$SAX spectra
 of BL Lac in the 0.3$-$100~keV  range using the following four 
models$^\dagger$: phabs*power, phabs*bbody, phabs*(bbody+power) and phabs*bmc.}
    \renewcommand{\arraystretch}{1.2}
    \label{tab:par_sax}
\tablehead{     & Parameter & S1 & S2 & S3 & S4 & S5 }
\startdata
Model &      &        &        &       &      &     \\
phabs       & N$_H$ ($\times 10^{21}$ cm$^{-2}$) & 5.7$\pm$0.5 & 2.5$\pm$0.9 & 6.8$\pm$0.09  & 7.6$\pm$0.2 & 2.9$\pm$0.6 \\
Power-law   & $\Gamma_{pow}$    & 2.01$\pm$0.09 & 1.89$\pm$0.05 & 1.71$\pm$0.04 & 1.94$\pm$0.09 & 2.61$\pm$0.04  \\
            & N$_{pow}^{\dagger\dagger}$ & 440$\pm$90 & 206$\pm$20 & 302$\pm$30 & 210$\pm$55 & 1686$\pm$110 \\
      \hline
           & $\chi^2$ {\footnotesize (d.o.f.)} & 1.15 (80)     & 1.29 (82)     & 1.3 (79) & 0.72 (79)   & 1.05 (116) \\
      \hline
phabs      & N$_H$ ($\times 10^{21}$ cm$^{-2}$) & 3.7$\pm$0.2 & 2.5$\pm$0.3 & 6.7$\pm$0.5  & 6.8$\pm$0.3 & 2.9$\pm$0.08 \\
Bbody      & T$_{BB}$  (keV)   & 1.00$\pm$0.02   & 0.84$\pm$0.01  & 1.19$\pm$0.01    & 1.00$\pm$0.01  & 0.67$\pm$0.01 \\
           & N$_{BB}^{\dagger\dagger}$ & 13.8$\pm$0.7 & 8.7$\pm$0.2 & 15.4$\pm$0.3 & 7.5$\pm$0.3 & 32.9$\pm$0.6 \\
      \hline
           & $\chi^2$ {\footnotesize (d.o.f.)} &  3.76 (80) & 12.5 (82) & 2.31 (79)& 3.06 (79)& 13.44 (116) \\
      \hline
phabs      & N$_H$ ($\times 10^{21}$ cm$^{-2}$) & 2.7$\pm$0.3 & 2.7$\pm$0.2 & 2.7$\pm$0.6  & 4.7$\pm$0.4 & 3.0$\pm$0.1 \\
Bbody      & T$_{BB}$ (keV)  & 0.8$\pm$0.1 & 0.38$\pm$0.03  & 1.02$\pm$0.03  & 0.64$\pm$0.16  & 0.37$\pm$0.05     \\
           & N$_{BB}^{\dagger\dagger}$ & 6.4$\pm$1.7  & 3.05$\pm$0.03   & 3.8$\pm$0.3   & 1.4$\pm$0.1 & 8.7$\pm$1.2  \\
Power-law  & $\Gamma_{pow}$    & 1.67$\pm$0.09 & 1.65$\pm$0.08 & 1.46$\pm$0.03  & 1.6$\pm$0.3 & 2.4$\pm$0.1 \\
           & N$_{pow}^{\dagger\dagger}$ & 300$\pm$100& 134$\pm$36 & 150$\pm$10   & 108$\pm$80 & 1290$\pm$220 \\
      \hline
           & $\chi^2$ {\footnotesize (d.o.f.)}& 1.34 (78) & 0.87 (80) & 1.78 (77) & 0.75 (77)& 0.97 (114) \\
      \hline
phabs      & N$_H$ ($\times 10^{21}$ cm$^{-2}$) & 2.1$\pm$0.7 & 1.5$\pm$0.6 & 6.9$\pm$0.1  & 7.4$\pm$0.3 & 2.9$\pm$0.1 \\
bmc        & $\Gamma_{bmc}$    & 1.8$\pm$0.1 & 2.07$\pm$0.09& 2.21$\pm$0.09   & 2.1$\pm$0.2    & 2.2$\pm$0.2  \\
           & T$_{s}$   (eV)    & 73$\pm$5       & 108$\pm$9    & 71$\pm$10       & 48$\pm$7       & 50$\pm$10    \\
           & logA$$            & -0.32$\pm$0.04  & 0.24$\pm$0.09& -0.72$\pm$0.09    & -0.86$\pm$0.04  & 0.35$\pm$0.07 \\
           & N$_{bmc}^{\dagger\dagger}$ & 0.36$\pm$0.09 & 0.6$\pm$0.2 & 0.98$\pm$0.07  & 2.5$\pm$0.3  & 3.08$\pm$0.06 \\
      \hline
           & $\chi^2$ {\footnotesize (d.o.f.)}& 1.05 (78) & 0.87 (80) & 1.17 (77)& 0.95 (77)& 1.08 (114) \\
      \enddata
\\$^\dagger$     Errors are given at the 90\% confidence level. 
$^{\dagger\dagger}$ The normalization parameters of blackbody and bmc components are in units of $L^{soft}_{34}/d^2_{10}$ erg s$^{-1}$ kpc$^{-2}$, 
where $L^{soft}_{34}$ is the soft photon luminosity in units of $10^{34}$ erg s$^{-1}$, $d_{10}$ is the distance to the 
source in units of 10 kpc, and the power-law component is in units of 
keV$^{-1}$ cm$^{-2}$ s$^{-1}$ at 1 keV. 
$T_{BB}$ and $T_{s}$ are the temperatures of 
the blackbody and seed photon components, respectively (in keV and eV). 
$\Gamma_{pow}$ and $\Gamma_{bmc}$ are the indices of the { power law} 
and bmc, respectively. 
\end{deluxetable}

%
%

\begin{table*}
 \caption{Best-fit parameters  of the $ASCA$, $Suzaku$ and $Swift$ spectra of BL Lac in the 0.45$-$10~keV range 
using the 
phabs*bmc model$^{a}$ 
}
    \label{tab:par_swift}
 \centering 
 \begin{tabular}{lllllllllr}
 \hline\hline
Satellite &  Observational ID &  $\alpha=\Gamma-1$ & $T_s$ (eV) &  $N_{BMC}^b$ & $N_H^c$    & $\chi^2_{red}$ (d.o.f.) \\
 \hline                                             
 \hline
$ASCA$      & 15505000    & 0.32$\pm$0.04 & 100$\pm$7  &  2.43$\pm$0.10 & 0.27$\pm$0.09 & 1.02 (338) \\ 
            & 73088000    & 0.71$\pm$0.02 & 99$\pm$5 &  3.19$\pm$0.08 & 0.42$\pm$0.08 & 0.86 (131) \\ 
            & 77000000    & 0.68$\pm$0.01 & 91$\pm$2 &  3.57$\pm$0.09 & 0.29$\pm$0.09 & 1.14 (252) \\ 
 \hline
$Suzaku$    & 701073010   & 1.02$\pm$0.03 & 143$\pm$10 &   7.25$\pm$0.63  & 0.27$\pm$0.07 & 1.15 (818) \\ 
            & 707044010   & 0.92$\pm$0.02 & 128$\pm$9  &   11.51$\pm$0.05 & 0.23$\pm$0.09 & 1.06 (521) \\ 
 \hline
$Swift$     &   Band-A  &  0.95$\pm$0.11 & 190$\pm$10 &   3.3$\pm$0.2 &  0.27$^d$& 0.72 (326) \\ 
            &   Band-B  &  1.00$\pm$0.20 & 288$\pm$7  &   4.5$\pm$0.1 &  0.27$^d$& 0.94 (160) \\ 
            &   Band-C  &  1.04$\pm$0.05 & 91$\pm$6   &   6.3$\pm$0.3 &  0.27$^d$& 0.97 (551) \\ 
            &   Band-D  &  1.08$\pm$0.17 & 143$\pm$8  &   8.1$\pm$0.5 &  0.27$^d$& 0.96 (272) \\ 
 \hline                                             
 \end{tabular}
\\$^a$     Errors are given at the 90\% confidence level. 
$^{b}$ The normalization parameters of blackbody and bmc components are in units of 
$L^{soft}_{34}/d^2_{10}$ erg s$^{-1}$ kpc$^{-2}$, 
where $L^{soft}_{34}$ is the soft photon luminosity in units of $10^{34}$ erg s$^{-1}$, $d_{10}$ is the distance to the 
source in units of 10 kpc. 
$^c$ $N_H$ is the column density for the neutral absorber, {in units of $10^{22}$ cm$^{-2}$} (see details in the text).
$T_{s}$ is the seed photon temperature (in eV). 
$^d$ indicates that a parameter was fixed. 
 \end{table*}



\begin{deluxetable}{llllrllllr}
\tablewidth{0in}
\tabletypesize{\scriptsize}
    \tablecaption{Best-fit parameters  of the $RXTE$ spectra of BL Lac in the 3$-$100~keV range 
using the 
phabs*bmc model$^{a}$ during 1999 observations (MJD 51150 -- 51570, R2 set).}
    \renewcommand{\arraystretch}{1.2}
    \label{tab:par_rxte}
\tablehead{ Observational & MJD &  $\alpha=$  & $N_{BMC}^b$ & $\chi^2_{red}$  & Observational & MJD &  $\alpha=$  & $N_{BMC}^b$ & $\chi^2_{red}$ \\
ID            &     &  $\Gamma-1$ &             & (d.o.f.)        & ID            &     &  $\Gamma-1$ &             & (d.o.f.)                        }
\startdata
30213-09-03-00 & 51162.49 & 0.2(1)  &  1.04(8) & 0.84 (90) & 30255-01-11-00 & 51470.72 & 0.9(1)   &  1.8(2) & 0.91 (86) \\
40175-05-01-00 & 51179.29 & 0.50(6) &  2.19(7) & 0.84 (90) & 30255-01-12-00 & 51470.99 & 0.8(1)   &  3.8(3) & 0.93 (86) \\
40175-05-02-00 & 51221.33 & 0.47(7) &  1.47(6) & 0.79 (90) & 30255-01-13-00 & 51471.06 & 1.16(9)  & 14.3(4) & 0.76 (77) \\
40175-05-03-00 & 51228.34 & 0.57(7) &  2.12(8) & 1.09 (90) & 30255-01-13-01 & 51471.13 & 0.8(2)   &  3.4(2) & 0.93 (77) \\
40175-05-04-00 & 51235.14 & 0.57(8) &  1.91(7) & 0.84 (90) & 30255-01-13-02 & 51471.26 & 0.8(1)   &  3.5(4) & 0.79 (86) \\
40175-05-05-00 & 51243.57 & 0.19(8) &  0.69(7) & 0.79 (90) & 30255-01-14-00 & 51471.20 & 0.7(2)   &  2.9(1) & 0.76 (86) \\
40175-05-06-00 & 51251.20 & 1.23(8) &    19(1) & 0.82 (90) & 30255-01-14-01 & 51471.50 & 0.8(2)   &  3.2(2) & 0.77 (86) \\
40175-05-08-00 & 51263.70 & 0.8(1)  &  6.72(8) & 0.74 (90) & 30255-01-15-00 & 51471.67 & 0.9(1)   &  5.2(1) & 0.78 (84) \\
40175-05-09-00 & 51270.03 & 0.82(9) &  4.45(6) & 0.99 (90) & 30255-01-16-00 & 51471.74 & 1.1(1)   & 14.2(3) & 0.82 (86) \\
40175-05-11-00 & 51287.04 & 0.50(7) &  1.47(5) & 0.93 (86) & 30255-01-16-01 & 51471.80 & 1.08(9)  & 14.9(4) & 0.92 (86) \\
40175-05-12-00 & 51291.01 & 0.51(7) &  1.61(7) & 0.76 (86) & 30255-01-16-02 & 51471.87 & 1.1(1)   & 14.8(3) & 0.97 (86) \\
40175-05-13-00 & 51298.33 & 0.3(1)  &  0.86(7) & 0.78 (86) & 30255-01-16-03 & 51471.99 & 1.2(2)   & 16.8(5) & 0.73 (86) \\
40175-05-14-00 & 51305.03 & 1.19(7) &    18(1) & 0.79 (86) & 30255-01-17-00 & 51472.06 & 1.0(1)   & 10.9(2) & 1.10 (86) \\
40175-05-15-00 & 51312.54 & 1.12(9) &  14(1)   & 0.72 (86) & 30255-01-17-01 & 51472.39 & 0.7(2)   &  3.44(9) & 1.00 (86) \\
40175-05-16-00 & 51319.76 & 0.3(1)  &  0.53(4) & 0.82 (86) & 30255-01-18-00 & 51472.46 & 0.54(9)  &  2.2(1) & 0.72 (86) \\
40175-05-17-00 & 51326.06 & 0.87(9) &  3.50(8) & 0.74 (86) & 30255-01-18-01 & 51472.53 & 0.5(1)   &  2.1(2) & 0.73 (86) \\
40175-05-18-00 & 51333.78 & 0.9(1)  &  5.7(1)  & 0.76 (86) & 30255-01-19-00 & 51472.53 & 0.5(1)   &  1.8(1) & 0.96 (86) \\
40171-01-01-00 & 51336.74 & 0.20(9) &  0.75(3) & 0.86 (86) & 30255-01-19-01 & 51472.60 & 0.4(1)   &  1.5(1) & 0.76 (86) \\
40171-01-02-00 & 51337.31 & 0.20(9) &  0.93(3) & 0.77 (86) & 30255-01-20-00 & 51472.71 & 0.9(1)   &  5.9(1) & 0.86 (86) \\
40171-01-03-00 & 51337.81 & 0.26(8) &  0.72(8) & 0.79 (86) & 30255-01-21-00 & 51472.99 & 0.93(8) &  9.6(2) & 0.74 (86) \\
40171-01-04-00 & 51338.80 & 0.51(7) &  1.49(7) & 0.80 (86) & 30255-01-21-01 & 51473.00 & 0.83(9) &  5.1(2) & 0.73 (86) \\
40171-01-05-00 & 51339.87 & 0.51(7) &  1.19(3) & 0.76 (86) & 30255-01-21-02 & 51473.06 & 0.8(1)   &  4.86(8) & 0.79 (86) \\
40175-05-19-00 & 51340.52 & 0.3(1)  &  0.71(8) & 0.79 (86) & 30255-01-22-00 & 51473.25 & 0.37(9) &  1.57(5) & 1.15 (86) \\
40175-05-20-00 & 51348.23 & 0.13(9) &  0.59(6) & 0.79 (86) & 30255-01-24-00 & 51473.78 & 0.65(8) &  2.18(3) & 0.74 (86) \\
40175-05-21-00 & 51352.99 & 0.2(1)  &  0.65(7) & 0.76 (86) & 30255-01-25-00 & 51473.99 & 0.64(7) &  2.31(3) & 1.01 (86) \\
40175-05-22-00 & 51362.06 & 1.16(8) & 12.4(5)  & 0.74 (86) & 30255-01-26-00 & 51474.23 & 0.47(6) &  1.45(7) & 0.88 (86) \\
40175-05-22-01 & 51361.99 & 1.27(8) &    23(1) & 0.84 (86) & 30255-01-27-00 & 51474.50 & 0.46(7) &  1.23(8) & 0.84 (86) \\
40175-05-23-00 & 51368.51 & 0.91(9) &   7.4(1) & 0.74 (86) & 30255-01-28-00 & 51474.78 & 0.49(8) &  1.58(4) & 0.97 (86) \\
40175-05-25-00 & 51383.01 & 0.41(5) &  1.33(5) & 0.86 (86) & 40175-05-39-00 & 51480.38 & 0.35(8) &  0.97(7) & 0.72 (86) \\
40175-05-26-00 & 51389.67 & 0.32(5) &  1.04(4) & 0.84 (86) & 40175-05-40-00 & 51487.97 & 0.92(9) &  9.7(1) & 0.79 (86) \\
40175-05-27-01 & 51398.61 & 0.62(4) &  2.92(4) & 0.84 (86) & 40171-01-06-00 & 51517.11 & 0.50(9) &  2.28(6) & 0.74 (86) \\
40175-05-28-00 & 51404.01 & 0.60(8) &  2.08(6) & 0.78 (86) & 40171-01-07-00 & 51517.83 & 0.5(1)   &  1.60(7) & 0.78 (86) \\
40175-05-29-00 & 51410.01 & 1.2(2)  &    23(1) & 0.87 (86) & 40171-01-08-00 & 51518.18 & 0.5(1)   &  1.95(7) & 0.89 (86) \\
40175-05-30-00 & 51417.27 & 1.2(3)  &    72(5) & 0.87 (86) & 40171-01-09-00 & 51518.44 & 0.50(9) &  2.12(6) & 0.81 (86) \\
40175-05-31-00 & 51425.70 & 1.1(1)   & 12.9(3) & 1.11 (86) & 40171-01-10-00 & 51518.72 & 0.89(8) &  6.01(7) & 0.83 (86) \\
40175-05-32-00 & 51430.76 & 0.46(9) &  1.72(9) & 0.98 (86) & 40171-01-11-00 & 51519.22 & 0.61(7) &  2.57(9) & 0.79 (86) \\
40175-05-33-00 & 51438.88 & 0.46(9) &  0.90(8) & 0.73 (86) & 40171-01-12-00 & 51519.50 & 0.48(9) &  1.91(4) & 0.95 (86) \\
40175-05-34-00 & 51446.07 & 0.12(6) &  0.59(4) & 0.93 (86) & 40171-01-13-00 & 51519.85 & 0.49(9) &  2.20(3) & 0.81 (86) \\
40175-05-35-00 & 51453.20 & 0.59(7) &  2.45(8) & 0.85 (86) & 40171-01-14-00 & 51520.11 & 0.3(1)   &  1.17(2) & 0.96 (86) \\
40175-05-36-00 & 51459.87 & 0.85(8) &  5.54(9) & 0.74 (86) & 40171-01-15-00 & 51520.43 & 0.35(8) &  1.27(5) & 1.09 (86) \\
40175-05-37-00 & 51466.65 & 0.40(7) &  1.39(8) & 0.96 (86) & 40171-01-16-00 & 51520.95 & 0.22(9) &  0.71(4) & 0.83 (86) \\
30255-01-01-01 & 51468.01 & 0.2(1)   &  1.1(7) & 0.79 (90) & 40171-01-17-00 & 51521.35 & 0.15(7) &  0.96(4) & 0.72 (86) \\
30255-01-02-00 & 51468.52 & 0.5(1)   &  3.5(4) & 0.76 (90) & 40171-01-18-00 & 51522.09 & 0.46(7) &  2.08(7) & 0.80 (86) \\
30255-01-06-00 & 51462.14 & 1.1(2)   & 14.8(7) & 0.91 (86) & 40171-01-19-00 & 51522.77 & 0.26(5) &  1.23(7) & 0.86 (86) \\
30255-01-04-00 & 51469.00 & 0.8(1)   &  4.2(3) & 0.76 (77) & 40171-01-20-00 & 51522.52 & 0.86(9) &  9.99(9) & 0.98 (86) \\
30255-01-05-00G & 51469.07 & 0.8(2)  &  5.8(4) & 0.75 (79) & 40171-01-21-00 & 51523.10 & 0.49(8) &  1.12(8) & 0.85 (86) \\
30255-01-06-01 & 51469.20 & 1.1(2)   & 19.6(5) & 0.91 (86) & 40171-01-22-00 & 51523.35 & 0.40(6) &  1.67(6) & 0.89 (86) \\
30255-01-05-01G & 51469.27 & 0.5(1)  &  1.9(1) & 0.91 (86) & 40171-01-23-00 & 51523.88 & 0.42(7) &  1.78(7) & 0.74 (86) \\
30255-01-06-02 & 51469.53 & 1.1(1)   & 17.9(4) & 0.91 (86) & 40171-01-24-00 & 51524.29 & 0.44(8) &  1.56(6) & 0.82 (86) \\
30255-01-07-00 & 51469.72 & 0.9(2)   &  5.9(1) & 0.91 (86) & 40171-01-25-00 & 51524.88 & 0.43(7) &  1.51(5) & 0.76 (86) \\
30255-01-08-00 & 51469.93 & 0.9(2)   &  6.2(1) & 0.91 (86) & 40171-01-26-00 & 51525.08 & 0.43(8) &  1.83(6) & 0.84 (86) \\
30255-01-09-00 & 51470.20 & 1.13(2)  & 111(5)  & 0.91 (86) & 40171-01-27-00 & 51526.09 & 0.50(9) &  1.91(7) & 0.94 (86) \\
30255-01-10-00 & 51470.54 & 0.9(2)   &  5.8(1) & 0.91 (86) & 40171-01-28-00 & 51526.94 & 0.51(8) &  1.85(6) & 0.98 (86) \\
      \enddata
\\$^a$     Errors are given at the 90\% confidence level. 
$^{b}$ The normalization parameters of blackbody and bmc components are in units of 
$L^{soft}_{34}/d^2_{10}$ erg s$^{-1}$ kpc$^{-2}$, 
where $L^{soft}_{34}$ is the soft photon luminosity in units of $10^{34}$ erg s$^{-1}$, $d_{10}$ is the distance to the 
source in units of 10 kpc. 
$N_H$, the column density for the neutral absorber, was fixed at $2.6\times 10^{21}$ cm$^{-2}$, 
$T_{s}$, the seed photon temperature, was also fixed at 70 eV  (see details in the text). 
\end{deluxetable}


%
%

\begin{table*}
 \caption{Parameterizations for reference and target sources}
 \label{tab:parametrization_scal}
 \centering 
 \begin{tabular}{lcccccc}
 \hline\hline                        
  Reference source  &       $\cal A$ &     $\cal B$     &  $\cal D$  &    $x_{tr}$      & $\beta$  &  \\
      \hline
GX~339--4    RISE  2004 & 2.24$\pm$0.01 &  0.51$\pm$0.02  &  1.0 & 0.039$\pm$0.002   &   3.5  \\
4U~1543--37  DECAY 2005 & 2.15$\pm$0.06 &  0.63$\pm$0.07  &  1.0 & 0.049$\pm$0.001   &   0.6$\pm$0.1  \\
%
 \hline\hline                        
  Target source     &      $\cal A$     &    $\cal B$    &  $\cal  D$  &   $x_{tr} [\times 10^{-5}]$ & $\beta$ \\
      \hline
BL Lacertae & 2.17$\pm$0.09 & 0.62$\pm$0.07   & 1.0  &   9.58$\pm$0.06 & 0.51$\pm$0.06  \\
 \hline                                             
 \end{tabular}
 \end{table*}

%
%

\begin{deluxetable}{lllllc} 
\tablewidth{0in}
\tabletypesize{\scriptsize}
    \tablecaption{BH masses and distances.}
    \renewcommand{\arraystretch}{1.2}
 \label{tab:par_scal}
\tablehead{   Source   & M$^a_{dyn}$ (M$_{\odot})$ & i$_{orb}^a$ (deg) & d$^b$ (kpc)  & $M_{fund.plane}$ (M$_{\odot}$) &M$_{scal}$ (M$_{\odot}$) }
\startdata
GX~339--4     &   $> 6^{(1)}$        &     ...       &  7.5$\pm$1.6$^{(2)}$      &...&   5.7$\pm$0.8$^c$ \\
4U~1543--47   &   9.4$\pm$1.0$^{(3, 4)}$ &  20.7$\pm$1.5$^{(5)}$ &  7.5$\pm$1.0$^{(3)}$, 9.1$\pm$1.1$^{(5)}$    &...&   9.4$\pm$1.4$^c$ \\
BL Lacertae$^{(6, 7, 8, 9)}$  & ... &     ...      & $\sim$300$\times 10^3$ & $\sim$1.7$\times 10^8$& $\ge 3\times10^{7}$ \\
      \enddata
\\References: 
(1) Mu$\grave n$oz-Darias et al. (2008); 
(2) Hynes et al. (2004);
(3) Orosz  (2003); 
(4) Park et al. (2004);
(5) Orosz et al. (1998);
(6) Urry et al. (2000);
(7) Ryle (2008);
(8) Liang \& Liu (2003);
(9) Oke \& Gumm (1974).
\\$^a$ Dynamically determined BH mass and system inclination angle, $^b$ Source distance found in the literature, 
$^c$ Scaling value found by ST09. 
\end{deluxetable}

\end{document}